# Physics-informed discrete element modeling for the bandgap engineering of cylinder chains

**Yeongtae Jang**

Division of Mechanical system engineering, Jeonbuk National University, Jeonju, 54896, Republic of Korea

Department of Mechanical Engineering, Pohang University of Science and Technology (POSTECH), Pohang 37673, Republic of Korea

**Eunho Kim***

Division of Mechanical system engineering, Jeonbuk National University, Jeonju, 54896, Republic of Korea

Advanced Transportation Machinery Research Center & LANL-CBNU Engineering Institute-Korea, Jeonbuk National University, Jeonju 54896, Korea

**Jinkyu Yang****

William E. Boeing Department of Aeronautics & Astronautics, University of Washington Seattle, WA 98195-2400, USA

Department of Mechanical Engineering, Seoul National University, Seoul 08826, Republic of Korea

**Junsuk Rho**

Department of Mechanical Engineering, Pohang University of Science and Technology (POSTECH), Pohang 37673, Republic of Korea

Department of Chemical Engineering, Pohang University of Science and Technology (POSTECH), Pohang 37673, Republic of Korea

POSCO-POSTECH-RIST Convergence Research Center for Flat Optics and Metaphotonics, Pohang 37673, Republic of Korea

## Abstract

We propose an efficient method to build a simple discrete element model (DEM) that accurately simulates the oscillation of a continuum beam. The DEM is based on the Timoshenko beam theory of slender cylindrical members and their corresponding wave dynamics in assembly. This physics-informed DEM accounts for multiple vibration modes of the constituting beam elements in wide frequency ranges. We construct various DEMs mimicking cylinder chains and compare their wave dynamics with those measured in experiments to validate the proposed method. Furthermore, we construct a graded woodpile chain of slender cylinders. We experimentally and numerically investigate the frequency bandgaps of the system and demonstrate the possibility of constructing a wide bandgap by consecutively superposing multiple stop bands generated from cylinders of various lengths. This system is highly efficient in blocking propagating waves by leveraging the vibration isolation effect stemming from the local resonance of the cylinders. The proposed DEM method can be useful for investigating and designing complex vibration systems in an efficient and accurate manner. Moreover, the design approach of manipulating the frequency bandgap can be exploited for developing vibration filters and impact mitigation devices.

* Corresponding Author
E-mail: eunhokim@jbnu.ac.kr
** Co-corresponding Author
E-mail: jkyang11@snu.ac.kr

# 1 Introduction

Phononic crystals (PCs) are structures consisting of periodically arranged vibrating elements that demonstrate wave phenomena depending on their periodicity. As a result, it is possible to design materials or structures with specific wave dynamics by carefully selecting the vibrating elements or controlling their periodic arrangement. The research field dedicated to designing such structures, known as acoustic/elastic metamaterials, has garnered significant interest since Liu et al. [1] experimentally demonstrated a sonic material. Numerous subsequent studies have been conducted, showcasing the immense potential of these structures for various engineering applications, such as vibration filters [2,3], acoustic or elastic guiding [4–6], impact mitigation [7–10], and seismic isolators [11,12], among others.

Depending on the style of construction, the metamaterials can be categorized into two types: continuum and discrete forms. The continuum form [13], which includes blocks, plates, and beams, possesses an infinite number of vibration modes and allows waves to propagate through the continuous deformation of the medium. In contrast, discrete forms [14,15], such as rigid particle packs or chains, usually possess finite number of vibration modes depending on the number of particles and transmit energy through interactions among discrete particles. It should be noted that in the discrete structure, the particle itself is a continuum. However, if the particle's resonance frequencies are sufficiently high and fall outside the frequency range of interest, it can be treated as a rigid discrete particle. In such a case, the interaction between particles usually relies on their contact behaviors [16]. When the contact exhibits inherent nonlinearity (e.g., Hertzian contact), the assembled structure can exhibit abundant wave dynamics ranging from linear to highly nonlinear regimes [14,17].

Among the different phononic crystal architecture types, interestingly, a woodpile structure, i.e., a stack of slender beams [3,9,18], possesses the characteristics of both discrete and continuum structures; it can be regarded as a discrete structure consisting of continuum particles having multiple vibration modes. Consequently, the numerous bending vibration modes exhibited by the slender beams interact with propagating waves through contacts, serving as local resonances within the discrete system. This local resonance leads to the formation of a resonance-induced frequency bandgap in the periodic chain, which effectively blocks propagating waves within the frequency range. Generally, each different local resonance gives rise to an individual frequency bandgap within the periodic chain. This is well explained and demonstrated in the previous research [3].

It was demonstrated that the local resonances of a beam in a discrete system are highly useful for manipulating both linear [3] and nonlinear [8,9,18] waves propagating through woodpile structures. The woodpile architecture of stacked cylinders can exhibit various wave dynamics, even in the absence of local resonances, through the manipulation of the contact angle between cylinders. In the linear wave regime, we can demonstrate tunable frequency bandgaps in diatomic arrangements [19], topological wave states [20], *in situ* topological band transitions [21], and boomerang waves with gradual variations in the contact angles [6]. In the nonlinear wave regime, we can demonstrate nonlinear resonances (nonlinear solitary waves) and antiresonances (dispersive waves) [22], the manipulation of the solitary wave speed [5], and asymmetric wave transmissions akin to a diode effect [6]. The local resonance allows the cylinder chain to realize more diverse waves, such as modulated nonlinear waves, nanoptera [18], highly efficient impulse dispersions [9] in the nonlinear regime, and the creation of a series of frequency passbands and bandgaps [3] in the linear regime.

Despite such discovery of rich wave dynamics, methods for the efficient manipulation of frequency band structures in the cylindrical chains remain elusive. This is due to the challenges in computing the interwoven dynamics of continuous cylindrical elements and their discrete assemblies in an accurate yet simple manner. In this study, we develop an efficient method for constructing a discrete element model (DEM) that accurately computes wave propagation in discrete architectures composed of continuum units. Using this DEM model, we efficiently analyze frequency bandgaps of various cylinder chains in detail. Furthermore, we demonstrate the feasibility of bandgap engineering by building a graded cylindrical chain and creating a broad frequency bandgap in it.

The discrete element model (DEM) has been widely used in condensed matter physics and wave dynamics of granule systems because it is simple but represents the waves in a complex system well. However, creating a DEM for a continuum structure with multiple eigenmodes is challenging. Thus, only a limited number of relevant studies have been performed. Matlack et al. [23] introduce an approach to designing continuum metamaterials with weak coupling between unit cells (called 'perturbative metamaterials') by mapping DEMs with desired features using an optimization method and finite element (FE) simulations. This enables a systematic inverse design of the metamaterials. Kim and Yang [3] developed a DEM corresponding to a cylindrical beam using an optimization method – to make the DEM satisfy two cut-off frequencies of the cylinder chain – which is limited to considering only two local resonance modes.

As mentioned in these two previous studies [3,23], an optimization method is commonly utilized to create mathematical models for complex systems when there is insufficient information to determine unknown parameters. It is worth noting that recently, there have been significant advancements in machine learning-based approaches, which have been successfully applied to various solid mechanics problems [24–26]. This emerging approach also seems promising for the construction of simplified models in our field as well. However, unlike these approaches, the DEM construction method proposed in this study takes an analytical approach, utilizing enough physical information from the analytic beam theory (Timoshenko beam theory) of the constituent continuum elements and the corresponding wave dynamics when assembled into a phononic crystal. This analytical approach allows the DEM to be extended virtually without limits on the number of eigenmodes while maintaining high accuracy in mimicking the bending vibrations of an arbitrary shaped beam. Thus, the proposed discrete element modeling method holds significant potential for studying diverse wave dynamics in complex systems characterized by local resonance. It also offers a valuable tool for analyzing and designing intricate metamaterials that integrate continuum and granule structures. Furthermore, this method offers substantial improvements in the efficiency of designing dynamic systems when utilized alongside various optimization techniques.

The remaining sections of this paper are organized as follows. Section 2 provides a detailed description of the cylinder chain under investigation and outlines the experimental method employed to measure the wave dynamics within the chain. Section 3 presents the analytical approach used to construct a discrete element model (DEM) that accurately simulates the oscillation of a continuum beam. Section 4 introduces the numerical analysis methods utilized, which include both the DEM and FEM (finite element method), to compute the wave dynamics within the cylinder chain. In Section 5, we validate the proposed DEM construction method and discuss the obtained results, focusing on the emergence of infinite bandgaps in the graded chain. Finally, in Section 6, we conclude the paper by summarizing the key findings and contributions of this study.

## 2 Experiments

In this section, we describe the cylindrical chain model and the experimental setup employed to measure the wave dynamics of the woodpile architecture. Fig. 1(a) present an image of the experimental setup consisting of a graded single-column woodpile structure, supporting jigs, and measurement systems. Furthermore, Fig. 1(b) depicts the sensor and actuator locations within the test setup, along with the signal acquisition procedure diagram. Each cylinder in the chain structure orthogonally makes contact at its center with the adjacent cylinders. The cylinders are made of fused quartz (Density $\rho$ =2187 $\mathrm{kg/m^3}$, Young's modulus $E = 72$ GPa, Poisson's ratio $v = 0.17$) with low material damping. Thus, we can more clearly observe the wave dynamics in the cylinder chain with minimum wave dissipation. We stack 26 cylinders for the chain, keeping the diameter of the cylinder 5 mm. The length of the cylinder gradually varies from 100 mm (at the top) to 50 mm (at the bottom) with a 2 mm length change. We apply a static pre-compression force (18 N) to the chain by placing a balance weight on top of the actuator. As the magnitude of the applied static force is significantly greater than the dynamic perturbation caused by the actuator, the wave dynamics can be approximated as linear regimes [3,19,27,28].

The vertically aligned cylinder chain, inherently unstable and prone to buckling under compressive loading, poses challenges for accurate measurements of propagating waves. To address this issue, we employ guide rods and very soft polyurethane foam to support the unstable single-column chain, as shown in Fig. 1(a). The guide rods facilitate vertical motion of the cylinders without hindering wave propagation along the vertical direction of the structure. Their primary function is to restrict the horizontal motion of the cylinders, ensuring structural stability. Additionally, the soft polyurethane foam, located at both sides of the cylinder, effectively mitigates buckling while minimally interfering with wave propagation. This is attributed to the foam's significantly lower stiffness compared to the cylinders, as detailed in the supplementary material [18].

We then introduce a small perturbation to the top cylinder of the chain structure using a piezoelectric actuator (Piezomechanik PSt 500\10\25). The actuator is driven by a ±5 V chirp signal (1Hz ~ 50kHz) generated by a function generator (Agilent 33210A) for a duration of 0.5 secs. To measure the contact force applied to the cylinder chain accurately, we attach a small piezoelectric ceramic disc (with a scale factor of 4.32 N/V) to the actuator's tip, as depicted in the inset of Figure 1(a). This ceramic disc establishes direct contact with the top cylinder, enabling precise measurements of the force. Meanwhile, a bottom force sensor (PCB 208 C01) captures the wave transmitted through the chain. Both the input force (measured at the top) and the output force (measured at the bottom) signals are recorded by an oscilloscope (InfiniiVision DSOX2014) with a sampling frequency of 100 kHz. Subsequently, these signals are transferred to a computer, where we employ a MATLAB program to calculate the frequency response function of the chain structure based on the force ratio (output force to input force).

In order to investigate the wave propagation inside the chain, we measure the velocity of each cylinder near the contact location. To achieve this, we affix pieces of reflection tape near the center of each cylinder. The velocity of each cylinder is then measured using Laser Doppler Vibrometer (LDV, Polytec OFV-5000/OFV-505) mounted on a linear guide as shown in Fig. 1(b). The velocity signal is also recorded on the oscilloscope and then transferred to the computer for the analysis.

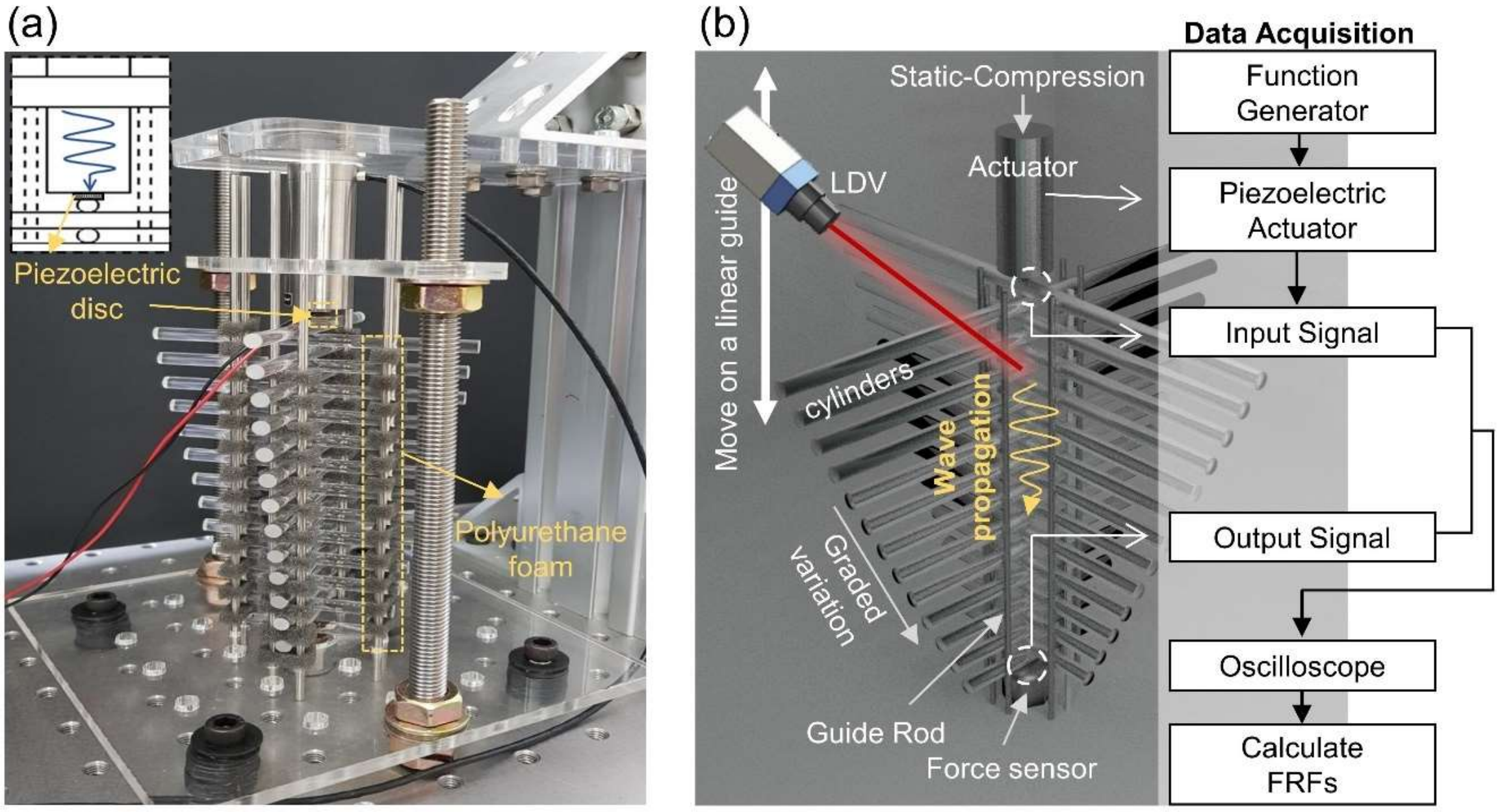

**Figure 1.** (a) Experiment setup for the one-dimensional graded cylinder chain. (b) A schematic diagram illustrating the signal acquisition procedure and the arrangement of the chain structure with an actuator and force sensors.

## 3 Analytical model

In this section, we explain the proposed analytical method for constructing a discrete element model (DEM) that accurately simulates the oscillations of a continuum beam. This method is built upon a combination of the continuum beam theory and the wave dynamics of a periodic structure.

Figure 2 presents a concise flowchart outlining the overall procedure, which will be further elaborated in subsequent sub-sections. Initially, we consider a periodic chain of beams (see (a) Periodic system). Within this chain, a single beam is regarded as a continuum unit cell (see (b) continuum model), having contact boundary conditions at the center and free boundary conditions at the edges. By employing beam theory with specified boundary conditions (BCs), we can analyze the vibration characteristics of the continuum unit cell (beam). The contact boundary condition (BC) at the cylinder center depends on the relative motions with neighboring cylinders, which are influenced by the wavenumber of the propagating wave. We can identify three types of characteristic frequencies by considering three specific BCs corresponding to different wave states.

On the other hand, in the discrete element model (c), the beam's bending vibration mode inducing local resonance is considered using a mass-with-mass model, where a spring-mass unit is attached to a primary mass ($\widetilde{M}$). To consider $N$ vibration modes in the discrete element method (DEM), $N$ spring-mass units can be added to the primary mass. From the DEM's equations of motion, we derive the dispersion relationships, i.e., the wavenumber and frequency relationships. By applying the characteristic frequencies obtained from the continuum beam theory to the dispersion relationships and solving the system of equations, we can analytically determine the masses and spring stiffness values of the DEM.

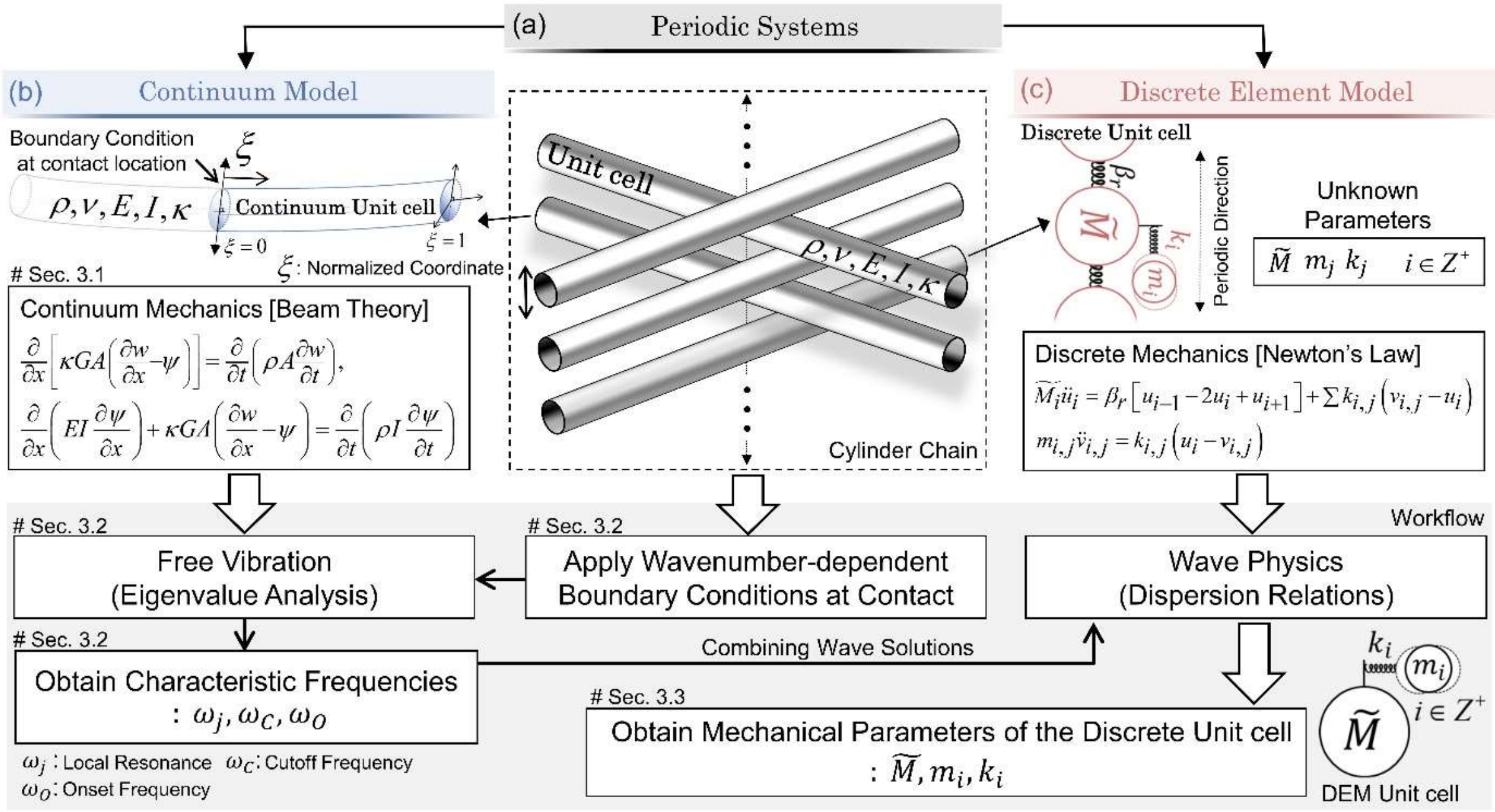

**Figure 2.** Flowchart depicting the construction of a DEM using the proposed physics-informed analytical method for simulating beam oscillations. (a) Periodic system representing a chain of cylinders serving as an initial assumption. (b) Continuum Model representing a continuum unit cell with boundary conditions, enabling the application of beam theory. (c) Discrete Element Model representing a DEM chain consisting of mass-with-mass unit cells and their corresponding equations of motion. The bottom diagram illustrates that the parameters of the DEM are determined by combining the characteristic frequencies obtained from the continuum beam theory and the dispersion relationships derived from the DEM.

### 3.1 Timoshenko beam theory

First, we assume a periodic cylinder chain as shown in Fig. 2(a). To analyze the vibration of the cylinder (continuum unit cell in Fig.2(b)) we use the Timoshenko beam theory, which can take into account the shear deformation and rotary inertia effect of a beam [29]. Typically, the Euler beam theory is employed to analyze the deformation of slender beams. However, in higher-order vibration modes, even slender beams exhibit complex deformations where shear deformation becomes significant. Therefore, utilizing the Timoshenko beam theory improves the accuracy by accounting for shear deformation. In the absence of an external force, the Timoshenko beam equation is expressed as a coupled partial differential, as follows:

$$\frac{\partial}{\partial x}\left[\kappa GA\left(\frac{\partial w}{\partial x}-\psi\right)\right]=\frac{\partial}{\partial t}\left(\rho A\frac{\partial w}{\partial t}\right), \tag{1}$$

$$\frac{\partial}{\partial x}\left(EI\frac{\partial \psi}{\partial x}\right)+\kappa GA\left(\frac{\partial w}{\partial x}-\psi\right)=\frac{\partial}{\partial t}\left(\rho I\frac{\partial \psi}{\partial t}\right). \tag{2}$$

Here $w(x,t)$ and $\psi(x,t)$ are the vertical displacement and rotation angle of a beam cross-section, respectively. Also, $\rho, E,$ and $G$ are density, Young's modulus, and shear modulus, respectively. In the isotropic material, $G=E/2(1+v)$, where $v$ is Poisson's ratio. $A$ and $I$ are the cross-sectional area and its moment of inertia, respectively. $\kappa$ is the shear correction factor which is $\kappa=6(1+v)/(7+6v)$ for a circular cross-section of the cylinder [30]. For a linear elastic, isotropic, and homogeneous beam with a constant cross-section, these two equations can be combined, as follows:

$$\underbrace{EI\frac{\partial^4 w}{\partial x^4}+\rho A\frac{\partial^2 w}{\partial t^2}}_{Euler-Bernoulli\,Beam\,Theory}\underbrace{-\rho I\left(1+\frac{E}{\kappa G}\right)\frac{\partial^4 w}{\partial x^2\partial t^2}+\frac{\rho^2 I}{\kappa G}\frac{\partial^4 w}{\partial t^4}}_{Shear\,deformation-Rotary\,inertia\,effect}=0. \tag{3}$$

This governing equation of the Timoshenko beam can be divided into the Euler-Bernoulli beam component and the considerations for shear deformation and rotary inertia effect of the beam. Under the harmonic motion, we can take $w(x,t)=V(x)e^{i\omega t}$, $\psi(x,t)=U(x)e^{i\omega t}$ into Eq. (3), where $\omega$ is the angular frequency. Then Eq. (1) ~ (3) is expressed as follows:

$$V_{xx}-U_x+\frac{\rho\omega^2}{\kappa G}V=0, \tag{4}$$

$$EIU_{xx}+\kappa GAV_x+\left(\rho I\omega^2-\kappa GA\right)U=0, \tag{5}$$

$$EIV_{xxxx} + \rho I\omega^2(1+\frac{E}{\kappa G})V_{xx} + \left(\frac{\rho^2 I\omega^4}{\kappa G} - \rho A\omega^2\right)V = 0, \tag{6}$$

where the subscript $x$ denotes differentiation with respect to the corresponding arguments of $U(V)$. We adopt non-dimensional variables for convenience as follows:

$$\xi = \frac{x}{L},\ \Omega^2 = \frac{\rho A\omega^2 L^4}{EI},\ \Gamma^2 = \frac{EI}{\kappa GAL^2},\ \Lambda^2 = \frac{I}{AL^2},\ W(\xi) = \frac{V(x)}{L},\ \Psi(\xi) = \frac{U(x)}{L}. \tag{7}$$

Similar to the aforementioned interpretation, Γ and Λ indicate the shear deformation and rotary inertia effect, respectively. Using these non-dimensional variables, the Eqs. (4) ~ (6) can be rewritten as follow:

$$W_{\xi\xi} + \Omega^2\Gamma^2 W - L\Psi_{\xi} = 0,\ \left[0 \le \xi \le 1\right], \tag{8}$$

$$\Gamma^2 L\Psi_{\xi\xi} - L(1-\Omega^2\Gamma^2\Lambda^2)\Psi + W_{\xi} = 0,\ \left[0 \le \xi \le 1\right], \tag{9}$$

$$W_{\xi\xi\xi\xi} + \Omega^2\left(\Gamma^2 + \Lambda^2\right)W_{\xi\xi} + \Omega^2\left(\Omega^2\Gamma^2\Lambda^2 - 1\right)W = 0,\ \left[0 \le \xi \le 1\right]. \tag{10}$$

If we assume $W(\xi) = e^{i\varphi\xi}$, the characteristic equation of Eq. (10) becomes

$$\varphi^4 - \Omega^2\left(\Gamma^2 + \Lambda^2\right)\varphi^2 + \Omega^2\left(\Omega^2\Gamma^2\Lambda^2 - 1\right) = 0. \tag{11}$$

From this quadratic equation, we obtain four characteristic roots, as follows:

$$\varphi_{1,2,3,4} = \pm\Omega\sqrt{\frac{\left(\Gamma^2 + \Lambda^2\right) \pm \sqrt{\left(\Gamma^2 - \Lambda^2\right)^2 + 4/\Omega^2}}{2}}. \tag{12}$$

Thus, the general solution of $W(\xi)$ becomes Eq. (13), and $\Psi(\xi)$ becomes Eq. (14) from Eq. (9) and Eq. (13).

$$W(\xi) = C_1\cosh(\Omega\alpha\xi) + C_2\sinh(\Omega\alpha\xi) + C_3\cos(\Omega\beta\xi) + C_4\sin(\Omega\beta\xi), \tag{13}$$

$$\Psi(\xi) = C_1'\sinh(\Omega\alpha\xi) + C_2'\cosh(\Omega\alpha\xi) + C_3'\sin(\Omega\beta\xi) + C_4'\cos(\Omega\beta\xi). \tag{14}$$

In low-frequency ranges satisfying the inequality, $\left[\left(\Lambda^2-\Gamma^2\right)^2+4/\Omega^2\right]^{1/2}>\left(\Lambda^2+\Gamma^2\right)$ in Eq. (12), $\alpha$ and $\beta$ are expressed as follows:

$$\begin{Bmatrix}\alpha\\ \beta\end{Bmatrix}=\frac{1}{\sqrt{2}}\left\{\mp\left(\Lambda^2+\Gamma^2\right)+\left[\left(\Lambda^2-\Gamma^2\right)^2+\frac{4}{\Omega^2}\right]^{1/2}\right\}^{1/2}. \tag{15}$$

In the opposite case of inequality, we can substitute $\alpha$ with $i\alpha$. The constants in Eqs. (13) and (14) have the following relationships:

$$\begin{Bmatrix}C_1'\\ C_2'\end{Bmatrix}=\left(\frac{\Omega}{L}\right)\frac{\alpha^2+\Gamma^2}{\alpha}\begin{Bmatrix}C_1\\ C_2\end{Bmatrix}, \tag{16}$$

$$\begin{Bmatrix}C_3'\\ C_4'\end{Bmatrix}=\left(\frac{\Omega}{L}\right)\frac{\beta^2-\Gamma^2}{\beta}\begin{Bmatrix}-C_3\\ C_4\end{Bmatrix}, \tag{17}$$

where the undetermined constants $C_{1,2,3,4}$ can be found by imposing four boundary conditions on the beam. Once the boundary conditions are applied, we can obtain the natural frequencies (eigenvalue) and the corresponding mode shapes (eigenvector) of the beam. The boundary condition is implemented as a wavenumber-dependent condition, as depicted workflow in Fig. 2, and be further explained in detail in the next section. It should be noted if the shear deformation in the beam is negligibly small (e.g, when a slender beam vibrates with low-frequency vibration modes), the Timoshenko beam equation can be simplified as the Euler beam equation by substituting $\Gamma=\Lambda=0$ in Eq. (10) (see Appendix D).

### 3.2 Boundary conditions of a cylindrical beam in a cylinder chain

In this section, we elaborate on the boundary conditions of the cylinder (the continuum unit cell) within the periodic cylinder chain. The cylinders have free boundaries at both edges ($\xi=\pm 1$) and make contact with neighboring cylinders at the center ($\xi=0$). We explain the details of the contact behavior and contact stiffness between neighboring cylinders in section 3.3.

Apart from the contact stiffness between the cylinders, it is possible to impose various boundary conditions at the cylinder center ($\xi=0$) depending on the relative motion of the neighboring cylinder. Considering plane waves in an infinite chain consisting of identical cylinders, we get the dispersion curve (normalized wavenumber vs. frequency) of the chain as shown on the left in Fig. 3(a), where $k$ is a wavenumber, and $a$ is a lattice constant which is the diameter of the cylinder. Notably, multiple frequency passbands and bandgaps emerge due to the local resonances created by the bending vibration modes of the cylinders in the cylinder chain [3]. In the dispersion curve (Fig. 3(a)), when the wavenumber is zero ($ka = 0$) (infinite wavelength), all cylinders exhibit synchronized motion with the identical displacement. As a result, there are no interactions with adjacent cylinders (Fig. 3(b) top). This scenario can be represented by the equivalent lumped-mass model shown in Fig. 3(c) top, where all masses move in phase. Consequently, there is no constraint at the center of the cylinder ($\xi = 0$) in the vertical direction in this case. At the maximum normalized wavenumber ($ka = \pi$) corresponding to the shorted wavelength, all of the cylinders exhibit an out-of-phase motion with neighboring cylinders (see Figs. 3(b) and 3(c) middle). Consequently, in this scenario, it is appropriate to impose elastic support with the contact stiffness at the contact locations ($\xi = 0$). Finally, when $ka = -\infty i$ ($i$ is the imaginary number), which occurs within the frequency bandgap, the wave does not propagate. Instead, the cylinders only exhibit local vibrations without any displacement at the contact locations ($\xi = 0$). In this case, a fixed boundary condition can be imposed at the cylinder center ($\xi = 0$).

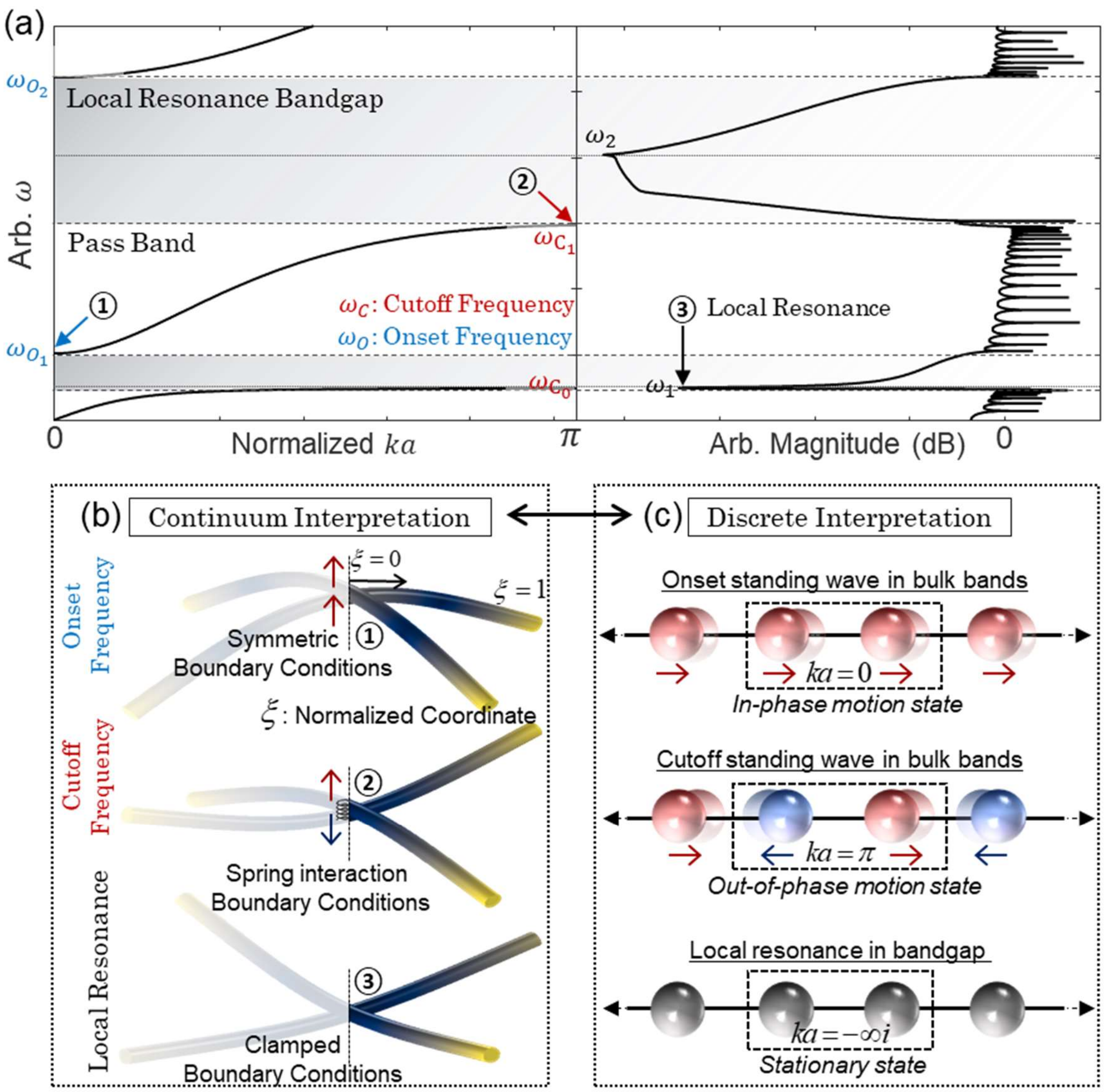

**Figure 3.** (a) The dispersion curve (left), normalized wavenumber vs. frequency, of a homogeneous cylinder chain and the corresponding frequency response function (right), frequency vs. magnitude, are presented in the same frequency range (vertical axis), where the shaded area indicates the local resonance bandgap. $\omega_{O_i}$ and $\omega_{C_i}$ are onset and cutoff frequency of each passband, respectively. $\omega_i$ is local resonance frequency. (b) Mode shapes of the cylindrical beam (local motion of the cylinders) at the frequency points labeled as ①, ②, and ③ on the dispersion curve (or FRF). These points correspond to distinct boundary conditions applied at the center ($\xi = 0$) of the beam (continuum unit cell): for onset frequencies at $\kappa a = 0$ – symmetric (sliding) BCs, for cutoff frequencies at $\kappa a = \pi$ – shear spring BCs, and for local resonance frequencies at $\kappa a = -\infty i$ – clamped BCs. (c) Schematics of standing wave states (global motion of particles) at three particular wavenumber states: $\kappa a = 0$ (corresponding to the onset frequency of the passbands), $\kappa a = \pi$ (corresponding to the cutoff frequency of the passbands),

and $\kappa a = -\infty i$ (corresponding to the local resonance frequency in the bandgaps). Here, we express the beam as a point mass to express the global motion.

In these three wavenumber states, the cylinder's natural frequencies and mode shapes can be determined by applying the respective boundary conditions. In this analysis, we consider only half of the cylinder due to its symmetry at its center ($\xi = 0$). Notably, the cylinder's asymmetric bending modes (even-numbered bending modes) do not influence the propagation of waves when the cylinders come into contact at their center (nodal points). Consequently, the boundary conditions for the three cases are as follows:

### *I. Wavenumber* $ka = 0$

When all cylinders move in phase without relative displacements at $ka = 0$, we can apply the sliding boundary condition at the cross section of the cylinder at $\xi = 0$ and apply the free boundary condition at $\xi = 1$ (Fig. 3. (c) top) as follows:

Sliding BCs: $\left.\Psi_{\xi}\right]_{\xi=0} = 0,\ \left.\kappa GA\left(W_{\xi} - L\Psi\right)\right]_{\xi=0} = 0,$ (18)

Free BCs: $\left.\Psi_{\xi}\right]_{\xi=1} = 0,\ \left.\kappa GA\left(W_{\xi} - L\Psi\right)\right]_{\xi=1} = 0.$ (19)

Then, Eqs. (13) & (14) can provide a characteristic equation. The frequencies satisfying the characteristic equation correspond to the onset frequencies ($\omega_{O_i}$) of each passband in the dispersion curve (Fig. 3(a)).

### *II. Wavenumber* $ka = \pi$

At $ka = \pi$ in the dispersion curve (Fig. 3(a) left), the neighboring cylinders move against each other (out-of-phase motion). In this state, the cylinder motion is predominantly influenced by the contact force between neighboring cylinders. To account for this, we connect two springs with linearized contact stiffness to each cylinder (one at the top and the other at the bottom). The deflection of these springs is twice the cylinder displacement resulting from the cylinders' out-of-phase motions. Assuming that the spring force applied at the center of the cylinder ($\xi = 0$) is equal to the shear force at the cylinder center, we can apply two boundary conditions at $\xi = 0$ in the half beam. One is a symmetric BC in terms of its cross-section surface, and the other is a shear force applied due to contact, expressed as follows:

$$\left.\Psi_{\xi}\right]_{\xi=0} = 0,\ \left.W_{\xi} - L\Psi\right]_{\xi=0} = \left.2\eta_r W\right]_{\xi=0}, \quad (20)$$

where $\eta_r$ is a dimensionless linearized contact stiffness expressed as $\beta_r L / \kappa GA$, and $\beta_r$ is a linearized contact stiffness between two cylinders which will be explained in section 3.3. The BC at $\xi = 1$ in the half beam is identical to that in Eq. (19). The characteristic frequencies are then calculated by Eqs. (13) & (14) with the given BCs. These correspond to the cutoff frequencies ($\omega_{c_i}$) of each passband as shown in the dispersion curve (Fig. 3(a) left) and the frequency response (Fig. 3(a) right).

#### *III. Wavenumber* $ka = -\infty i$

In the dispersion curve, a state $ka = -\infty i$ exists for each frequency bandgap due to local resonance (i.e., symmetric bending vibration mode of the beam), which corresponds to ideally zero magnitudes in the frequency response function (Fig. 3(a) right). In this state, waves cannot propagate, and the contact location ($\xi = 0$) in each cylinder remains stationary. Therefore, we can apply a fixed boundary condition at $\xi = 0$ in the half beam, as follows:

$$W\big]_{\xi=0} = 0,\ \Psi\big]_{\xi=0} = 0. \tag{21}$$

Also, we apply free BCs at $\xi = 1$ in the half beam (Eq. (19)). Then, the characteristic frequencies calculated with these BCs become the locally resonant frequencies ($\omega_i$) of each beam.

### 3.3 Discrete element model of a cylindrical beam

In this section, we describe a method by which to construct an equivalent DEM simulating the vibration of a continuum beam by using the characteristic frequencies calculated in the previous section. A cylindrical beam has multiple bending vibration modes, and each mode serves as a local resonance under dynamic excitation. The equivalent DEM precisely mimics the local resonance of the continuum beam by using spring-mass sets. To incorporate the local resonances effectively into the DEM, we employ a mass-with-mass type model, commonly used to represent local resonance phenomena in DEM [3,18]. Specifically, we connect multiple spring-mass ($m_j, k_j$) sets to a primary mass ($\widetilde{M}$). These unit cells are connected using springs ($\beta_r$), which represent the contact stiffness between cylinders, in the periodic chain model (see Fig. 4(a)).

The contact behavior between two cylinders is well described by nonlinear Hertzian contact, $\mathrm{F} = B_r\delta^{3/2}$; here $B_r$ is the contact coefficient defined as $B_r \equiv \frac{2E_r\sqrt{R_r}}{3(1-v_r{}^2)}$, where $E_r$, $v_r$, and $R_r$ are Young's modulus, Poisson's ratio, and the radius of the cylinder, respectively [16]. $\delta$ denotes the compression between two cylinders. When the dynamic force ($F_d$) is much weaker than the static pre-compression ($F_0$), i.e., ($F_d \ll F_0$), we can linearize the contact stiffness as follows:

$$\beta_r = \frac{3}{2}\sqrt[3]{B_r^2}\sqrt[3]{F_0}. \tag{22}$$

If we express the unit cell having multiple spring-mass sets (Fig. 4(a)) as a single effective mass (Fig. 4(b)), it can be expressed as a function of the frequency ($\omega$), as follows:

$$M_{eff}(\omega) = \widetilde{M} + \sum_{j=1}^{N} m_j \frac{\omega_j^2}{\omega_j^2 - \omega^2}, \tag{23}$$

where $\omega_j = \sqrt{k_j / m_j}$, $j \in [1, N]$ is the local resonance frequency of the $j^{\text{th}}$ spring mass set. The details of the effective mass are elaborated in Appendix A. Importantly, we can consider $N$ number of local vibration modes using the same number of spring-mass sets, and there is no limitation to the available number of local resonances.

Now we determine the mass ($\widetilde{M}, m_j$) and stiffness ($k_j$) of the unit cell for a DEM (Eq. (23)) that can accurately reproduce the same wave dynamics of a continuum cylinder within a cylinder chain. If we present the unit cell as an effective mass ($M_{eff}$), the equation of motion for the $i^{\text{th}}$ unit cell in the chain (Fig. 4(b)) can be expressed as follows:

$$\left( \widetilde{M} + \sum_{j=1}^{N} m_j \frac{\omega_j^2}{\omega_j^2 - \omega^2} \right) \ddot{u}_i = \beta_r \left( u_{i-1} - 2u_i + u_{i+1} \right). \tag{24}$$

When a wave propagates through a periodic structure, we can represent the displacement $u_i$ in the form $u_i = U_i e^{i(\omega t - kx)}$ and $u_{i\pm1} = u_i e^{\mp ika}$ on the basis of Floquet's theorem [31]. By substituting them into Eq. (24), we can derive the following dispersion relationship:

$$\left( \widetilde{M} + \sum_{j=1}^{N} m_j \frac{\omega_j^2}{\omega_j^2 - \omega^2} \right) \omega^2 = 2\beta_r \left[ 1 - \cos(ka) \right]. \tag{25}$$

This equation represents the relationship between the wavenumber and frequencies of the waves propagating through the periodic chain, which is expected to match that of the continuum cylinder chain. Hence, by substituting the characteristic frequencies derived from the continuum beam theory at specific wave numbers into Eq. (25), we can determine the masses and stiffnesses of the unit cell in the DEM.

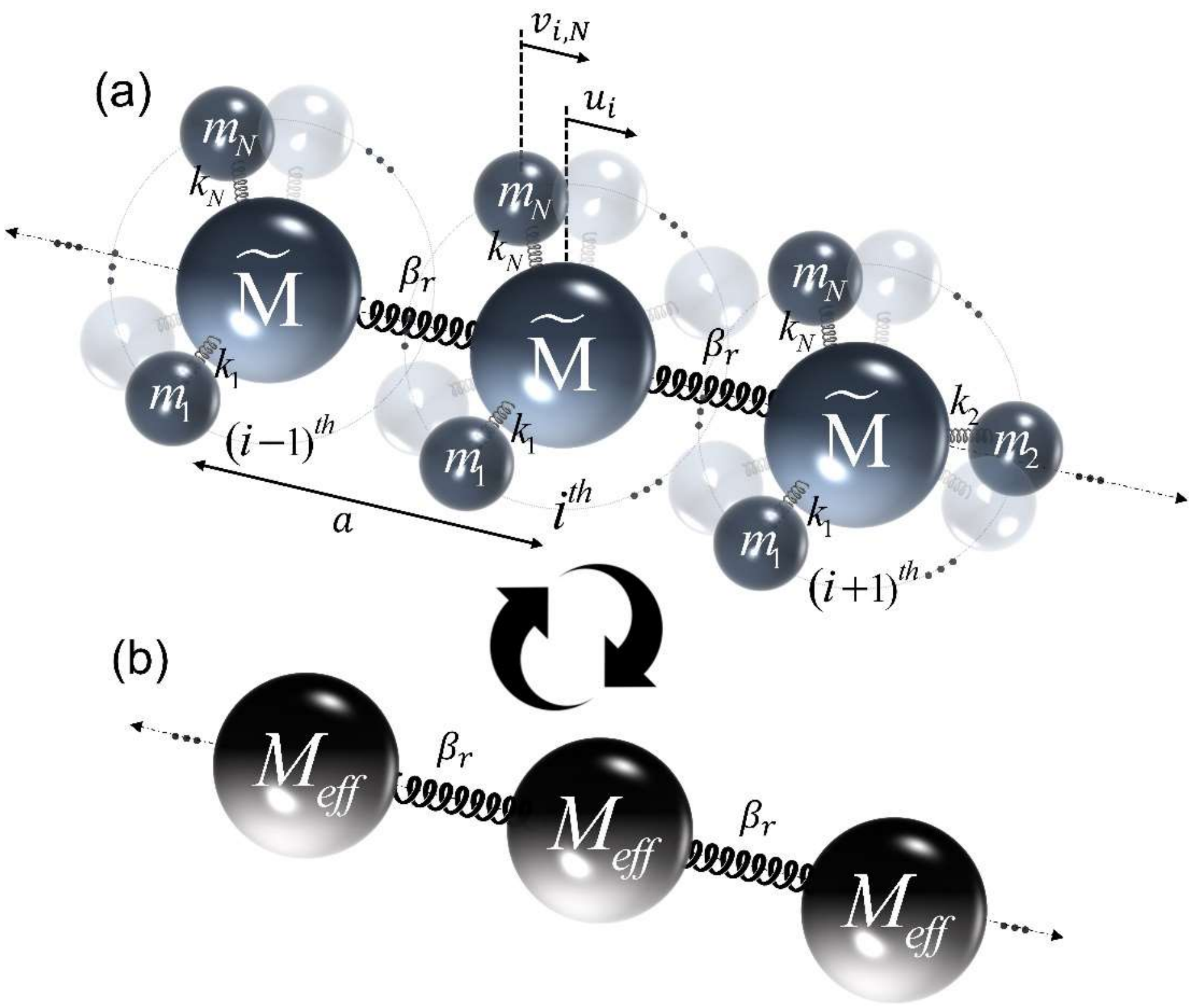

**Figure 4.** (a) Discrete element model (DEM) of a periodic chain incorporating a unit cell with multiple spring-mass sets to account for local resonance phenomena. (b) Corresponding DEM expressing the mass-with-mass unit cell as a single effective mass. In this representation, the effective mass is expressed as a function of the frequency. Each discrete particle (unit cell) is connected to linearized contact stiffness based on Hertzian theory.

The dispersion relationship is given by Eq. (26) for $ka = 0$ and by Eq. (27) for $ka = \pi$.

$$\left(\widetilde{M}+\sum_{j=1}^{N} m_j \frac{\omega_j^2}{\omega_j^2-\omega^2}\right)\omega_{O_l}^2 = 0, \quad l \in [0,\infty], \tag{26}$$

$$\left(\widetilde{M}+\sum_{j=1}^{N} m_j \frac{\omega_j^2}{\omega_j^2-\omega^2}\right)\omega_{C_l}^2 = 4\beta_r, \quad l \in [0,\infty], \tag{27}$$

where $\omega_{O_l}$ and $\omega_{C_l}$ are the onset frequency and cutoff frequency of the $l^{th}$ passband, respectively. It is noted that the index $l$ starts from zero (see Fig. 3(a) left). In the $0^{th}$ passband, the cylinders exhibit rigid body motion while interacting with neighboring particles. If we express $\omega^2$ as an eigenvalue $\lambda$ for simplicity, Eq. (27) for $l=0$ can be expressed as follows:

$$\left(\widetilde{M}+\sum_{j=1}^{N} m_j \frac{\lambda_j}{\lambda_j-\lambda_{C_0}}\right) = \frac{4\beta_r}{\lambda_{C_0}}. \tag{28}$$

Also, if we subtract Eq. (26) from Eq. (27), we get the following equation for $l \geq 1$,

$$\sum_{j=1}^{N} m_j \frac{\lambda_j \left(\lambda_{O_N} + \lambda_{C_N}\right)}{\left(\lambda_j - \lambda_{C_N}\right)\left(\lambda_j - \lambda_{O_N}\right)} = \frac{4\beta_r}{\lambda_{C_l}}. \tag{29}$$

If we consider a frequency range from zero to the $l^{th}$ passband, we get a total $(l+1)$ number of equations from Eq. (28) and Eq. (29). This is expressed in matrix form below.

$$\begin{bmatrix} 1 & \frac{\lambda_1}{\lambda_1 - \lambda_{C_0}} & \cdots\ \cdots & \frac{\lambda_N}{\lambda_N - \lambda_{C_0}} \\ 0 & \frac{\lambda_1(\lambda_{C_1} - \lambda_{O_1})}{(\lambda_1 - \lambda_{C_1})(\lambda_1 - \lambda_{O_1})} & \frac{\lambda_2(\lambda_{C_1} - \lambda_{O_1})}{(\lambda_2 - \lambda_{C_1})(\lambda_2 - \lambda_{O_1})} \cdots & \frac{\lambda_N(\lambda_{C_1} - \lambda_{O_1})}{(\lambda_N - \lambda_{C_1})(\lambda_N - \lambda_{O_1})} \\ \vdots & \frac{\lambda_1(\lambda_{C_2} - \lambda_{O_2})}{(\lambda_1 - \lambda_{C_2})(\lambda_1 - \lambda_{O_2})} & \frac{\lambda_2(\lambda_{C_2} - \lambda_{O_2})}{(\lambda_2 - \lambda_{C_2})(\lambda_2 - \lambda_{O_2})} \cdots & \frac{\lambda_N(\lambda_{C_2} - \lambda_{O_2})}{(\lambda_N - \lambda_{C_2})(\lambda_N - \lambda_{O_2})} \\ \vdots & \vdots & \vdots\ \ \ddots & \vdots \\ 0 & \frac{\lambda_1(\lambda_{C_l} - \lambda_{O_l})}{(\lambda_1 - \lambda_{C_l})(\lambda_1 - \lambda_{O_l})} & \frac{\lambda_2(\lambda_{C_l} - \lambda_{O_l})}{(\lambda_2 - \lambda_{C_l})(\lambda_2 - \lambda_{O_l})} \cdots & \frac{\lambda_N(\lambda_{C_l} - \lambda_{O_l})}{(\lambda_N - \lambda_{C_l})(\lambda_N - \lambda_{O_l})} \end{bmatrix} \begin{bmatrix} \widetilde{M} \\ m_1 \\ m_2 \\ \vdots \\ \vdots \\ m_{N-1} \\ m_N \end{bmatrix} = \begin{bmatrix} \frac{4\beta_r}{\lambda_{C_0}} \\ \frac{4\beta_r}{\lambda_{C_1}} \\ \vdots \\ \vdots \\ \frac{4\beta_r}{\lambda_{C_l}} \end{bmatrix}. \tag{30}$$

It should also be that the presence of $N$ number of local resonances in the mass-with-masses system leads to $N$ additional frequency passbands (excluding the $0^{th}$ passband). From Eq. (30), we can determine all masses $(\widetilde{M}, m_j)$ in the DEM unit cell. Then, the equation governing the locally resonant frequency, $\omega_j = \sqrt{k_j / m_j} \in [1, N]$, allows us to determine all of the spring stiffnesses $(k_j)$.

Once we obtain the DEM unit cell, we can simulate various wave dynamics of the cylinder chain spanning from linear to highly nonlinear regimes by introducing nonlinear springs between unit cells [6,9,18]. Here, we calculate the dispersion curve of the periodic chain within the linear wave regime using the following equation:

$$ka = \cos^{-1}\left[1 - \left(\widetilde{M} + \sum_{j=1}^{N} m_j \frac{\omega_j^2}{\omega_j^2 - \omega^2}\right)^2 \Big/ 2\beta_r\right]. \tag{31}$$

Figure 5 shows a dispersion curve calculated using a DEM chain, representing a periodic chain consisting of the longest quartz cylinders (100 mm) in the experimental setup. We consider ten local resonances (N = 10) in the DEM, which represents that the DEM can consider ten symmetric bending modes (i.e., ten odd-numbered bending modes) of the cylinder. It has been verified that the total mass of the unit cell $\left(\widetilde{M} + \sum_{j=1}^{N} m_j\right)$ asymptotically approaches the mass of the cylinder as the number of local resonances (N) increases. When employing ten local resonances (N = 10), the total mass is nearly identical to the continuum cylinder mass; the discrepancy is less than 0.00001%.

In the dispersion curve in Fig. 5, we can observe that the imaginary part (I.P.) drops unboundedly near each local resonance frequency (horizontal black dotted lines in I.P.), which means that the wave strongly attenuates in the vicinity of the local resonance frequencies. Interestingly, the width of the passband gradually narrows while the bandgap gradually becomes widen and stronger as the frequency increases. The DEMs constructed using the proposed method will be validated by comparing the wave dynamics in each DEM with those obtained from a finite element simulation and the experimental data in Section 5.

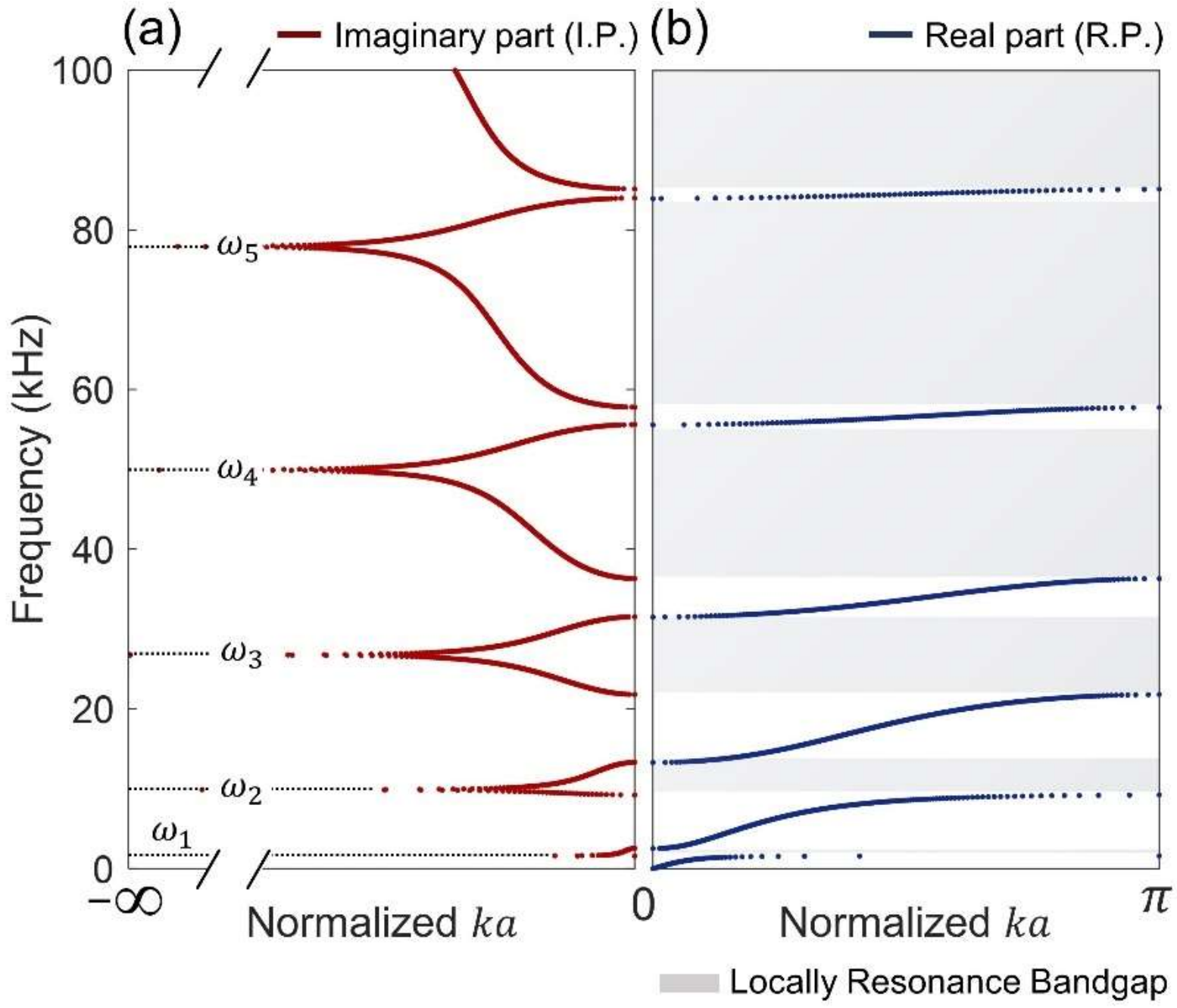

**Figure 5.** Dispersion curve calculated using a DEM corresponding to a periodic chain consisting of quartz cylinders with a length of 100 mm length and a diameter of 5 mm. It shows two parts: (a) the imaginary part (I.P.) and (b) the real part (R.P.) of the dispersion curve. The gray-shaded regions indicate the bandgap frequencies caused by local resonances

# 4 Numerical approach

In this section, we present numerical simulations of wave propagation in the finite graded cylinder chain presented in Fig.1. We use two numerical models: one is the DEM proposed herein, and the other is a finite element model (FEM).

## 4.1 Discrete element model

We first construct the DEMs for each cylinder employed in the experiment setup (Fig. 1), a total of 26 cylinders having different lengths from 100 mm to 50 mm with a 2 mm difference, in accordance with the method outlined in Section 3. Each DEM is equipped with ten local resonances, which sufficiently cover the frequency range of interest up to 50 kHz. The equations of motion of the $i^{th}$ cylinders are expressed as follows:

$$\widetilde{M_i}\ddot{u}_i = \beta_r\left[u_{i-1}-u_i\right]-\beta_r\left[u_i-u_{i+1}\right]+\sum_{j=1}^{N}k_{i,j}\left(v_{i,j}-u_i\right),\ \ i\in\{2,\ldots,P-1\},$$
$$m_{i,j}\ddot{v}_{i,j} = k_{i,j}\left(u_i-v_{i,j}\right),\ \ i\in\{2,\ldots,P-1\}, \tag{32}$$

where the subscript $(i,j)$ represents the $j^{th}$ local spring-mass unit in the $i^{th}$ cylinder model. $N$ represents the number of local modes considered and $P$ represents the total number of cylinders in the chain.

To account for the boundary condition of the finite chain, we modify the first equation in Eq. (32) for the first and the last particles as follows:

$$\widetilde{M_1}\ddot{u}_1 = F_d-\beta_w\left[u_1\right]-\beta_r\left[u_1-u_2\right]+\sum_{j=1}^{N}k_{1,j}\left(v_{1,j}-u_1\right),$$
$$\widetilde{M_P}\ddot{u}_P = -\beta_w\left[u_P\right]+\beta_r\left[u_{P-1}-u_P\right]+\sum_{j=1}^{N}k_{P,j}\left(v_{P,j}-u_P\right), \tag{33}$$

where $F_d$ represents dynamic perturbation applied to the first particle by the piezoelectric actuator used in the experiment. In this study, we apply a frequency sweep signal ranging from 1 Hz to 50 kHz to the perturbation $F_d$, with a magnitude of 0.01 [N]. It is important to note that this magnitude is significantly smaller than the static pre-compression force $F_0 = 18$ [N]. Additionally, the effect of gravity is disregarded, as its magnitude is orders of magnitude smaller compared to the static pre-compression force. In Eq.(33) $\beta_w$ represents the contact stiffness between the first(last) cylinder and the actuator tip (sensor tip), which is expressed as $\beta_w = {}^{3}/_{2}\, {B_w}^{2/3}{F_0}^{1/3}$, where $B_w \equiv \frac{2\sqrt{D_w B_1}}{3(V_w+V_c)}\cdot\frac{1}{B_2^{1.5}}$ is the contact coefficient [32]. $D_w$ is a diameter of the sensor cap curvature (0.6 m), $V_w \equiv \frac{1-v_w^2}{\pi E_w}$ and $V_c \equiv \frac{1-v_c^2}{\pi E_c}$, where $E_w(200\ GPa)$ and $v_w(0.29)$ are Young's modulus and Poisson's ratio of the boundary structure. We take $B_1 = 3.40$ and $B_2 = 4.45$ from our previous study [3].

To analyze the wave dynamics in finite cylinder chains using the DEM, we numerically solve the differential equations using the Runge-Kutta method (RK45) in MATLAB. For the steady-state response, we use the state-space model approach, which is further explained in Appendix B. We conduct an analysis of the wave dynamics in two types of cylinder chains: homogeneous cylinder chains and the graded cylinder chains. Firstly, we analyze several homogeneous chains consisting of 23 identical cylinders with the same length to understand the wave dynamics of a finite periodic structure and the effects of the cylinder length on the wave dynamics. Subsequently, we analyze the graded cylinder chain to investigate the superposition of bandgaps resulting from the gradual variation in the cylinder length.

### 4.2 Finite element model

We also simulate the wave dynamics in the cylinder chain using the finite element method (FEM, ABAQUS/STANDARD), as shown in Fig. 6. We utilize a three-node beam element (B32) implemented in the ABAQUS program for the modeling of the quartz cylinder. This beam element incorporates the Timoshenko beam theory, which accounts for transverse shear deformation. We confirm that the Timoshenko beam is more accurate than the Euler beam when we consider a high-frequency vibration mode of a beam, as the mode shape of a beam becomes more complex as its eigenfrequency increases, and the shear deformation and rotary inertia effect become more significant. To enhance the computational efficiency, we introduce axial springs with linearized contact stiffness ($\beta_r$) between adjacent cylinders, as shown in the inset of Fig. 6. By employing modal-based steady-state dynamic analysis, we obtain the frequency response function (FRF) of the cylinder chain up to 50 kHz. To ensure numerical stability, we incorporate a slight damping effect that accounts for global damping with a constant value of 0.0001 for all modes.

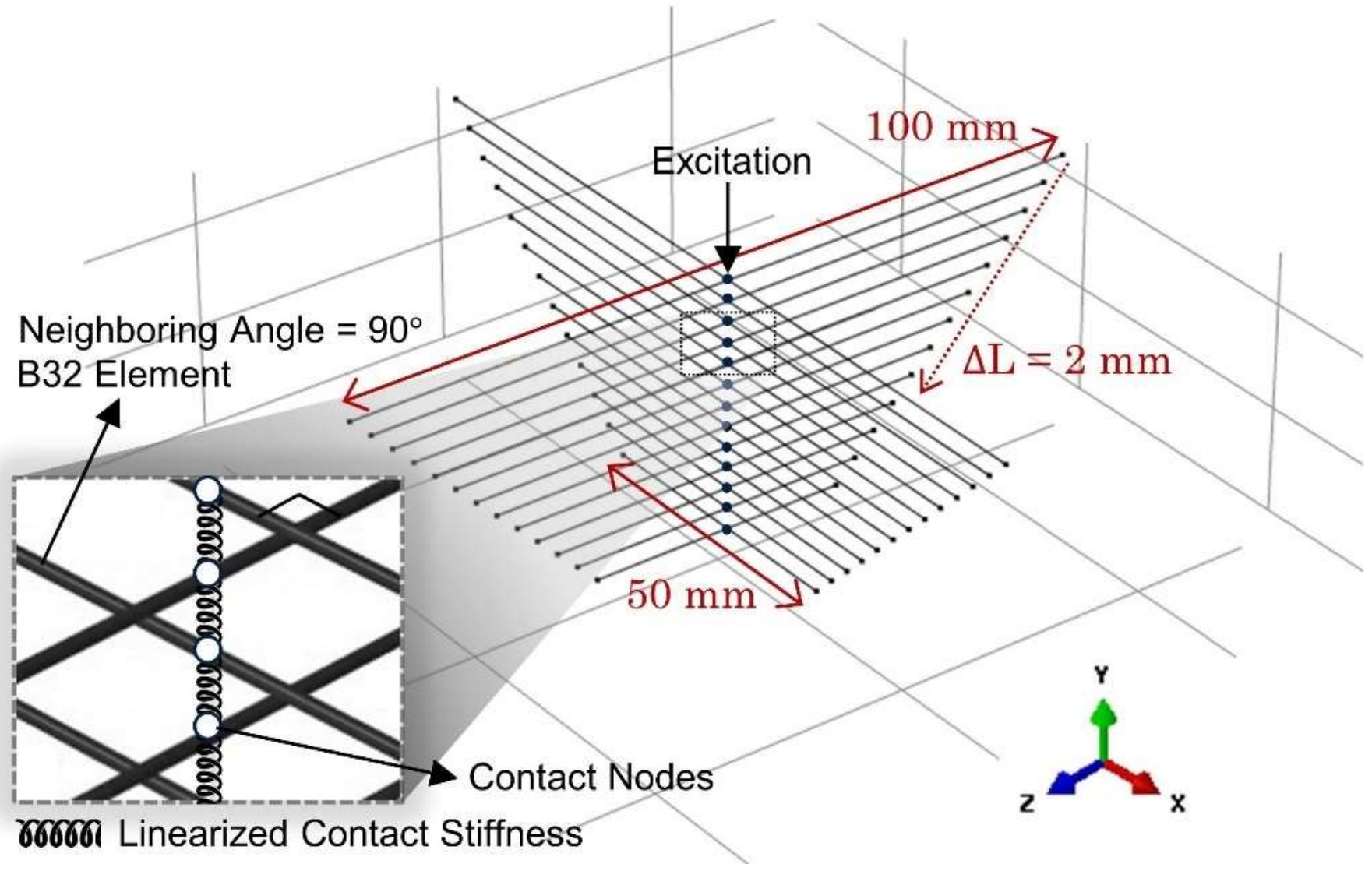

**Figure 6.** Finite element model of the cylinder chain. A three-node beam element (B32) supported in the ABAQUS/Standard program is used for the cylinder model. Each cylinder model is connected using a linear spring with a linearized contact stiffness.

# 5 Results and discussion

## 5.1 Frequency response of homogenous cylinder chains

Figure 7 shows frequency response functions (FRFs) of four representative cylinder chains with different cylinder length, 50 mm, 60 mm, 70 mm, and 80 mm, respectively. For each cylinder chain model, the experiment data measured up to 50 kHz are compared with two analysis data obtained from the DEM (see Appendix. B) and FEM. It is noted that the magnitude of each FRF obtained from DEM and FEM is scaled to fit the experiment data and they are shifted for easy comparison of the frequency band structure (passband and bandgap frequencies). The FRFs calculated using the DEM and FEM are in good agreement and they are comparable to the experiment data. In experiment, low frequency wave (the first passband) was not clearly captured. We presume that this is because the short time (0.5 secs) of chirp signal excitation is not enough to excite low frequency signals.

Each chain exhibits multiple sets of frequency passbands and bandgaps stemming from the coupled bending vibration modes of the cylinder interacting with propagating waves, creating local resonances within the system. The bending vibration modes of a cylinder repeatedly open and close the frequency bandgap, as explained in section 3. As the constituent cylinder length increases, the entire frequency band structure shifts towards lower frequencies (see the FRFs from bottom to top in Fig. 7). This occurs because the frequency band structure depends on the bending vibration mode of the cylinder, and the frequency of the mode decreases as the cylinder length ($L$) increases following the relationship of $\omega \propto {}^{1}/_{L^2}$.

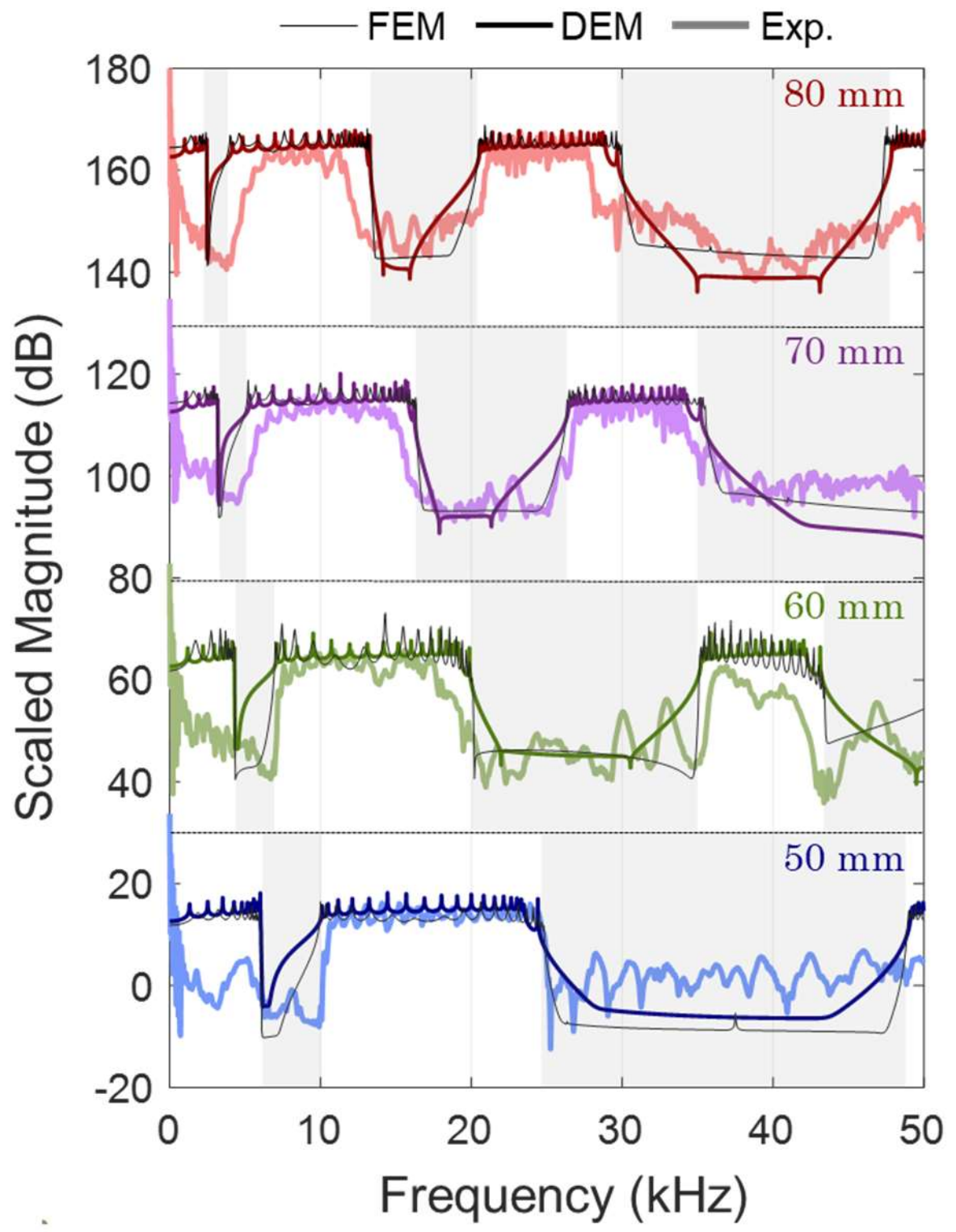

**Figure 7.** Comparison of frequency response functions (FRFs) of homogeneous cylinder chains consisting of 80 mm rods (first from the top), 70 mm rods (second), 60 mm rods (third), and 50 mm rods (the bottom). The thickest solid line with light color represents the FRF measured in the experiment [3], and the second thickest solid line with dark color represents the FRF calculated based on the DEM. The thinnest solid line with black color represents the FRF calculated based on the FEM. The FRF curves are scaled and shifted in a vertical direction for comparison. The shaded area represents the frequency bandgap.

We further investigate wave attenuation within the frequency bandgap in detail. First, we quantify wave attenuation in the bandgaps to determine the efficiency of wave attenuation. Assuming a harmonic wave within the homogeneous chains, the displacement of a particle can be expressed as $u_n = Ue^{i(ka(x/a)-\omega t)}$. If we introduce a complex form for the dimensionless wave number ($ka$), represented as $ka = \alpha + i\beta$, the displacement amplitude of the particle becomes proportional to $Ue^{-\beta(\frac{x}{a})}$, indicating that the wave amplitude exponentially decays with the coefficient $\beta$ in the spatial domain [33]. In other words, the imaginary part in the dispersion curve (associated with the bandgap frequency) signifies wave attenuation (see Fig. 5(a)). Notably, within the bandgap frequency range, the wave transforms into an evanescent wave, demonstrating total internal reflection. The degree of attenuation ($\beta$) varies with the frequency, as presented in Fig. 5(a). However, accurately discerning these variations in experimental data is challenging due to noise levels and experimental errors. Thus, we calculate the averaged attenuation coefficient ($\bar{\beta}$) in each bandgap frequency region to compare the efficiency of attenuation.

First, we experimentally (numerically) measure (calculated) the FRF for each cylinder in the homogeneous chains, after which we numerically integrate the amplitude of FRF in each bandgap frequency region ($\omega_{C_j} - \omega_{O_{j+1}}$) and divide it with the corresponding frequency range for each cylinder, which gives the average magnitude ( $\bar{\boldsymbol{M}}$ ) of FRF in the bandgap, $\bar{\boldsymbol{M}} = \int_{\omega_{C_j}}^{\omega_{O_{j+1}}} FRF(\omega)d\omega \big/ (\omega_{C_j} - \omega_{O_{j+1}})$. Finally, for each bandgap, we normalize $\bar{\boldsymbol{M}}$ of each cylinder based on the second cylinder's value, which is the largest value except for the first cylinder. We exclude the first one because it is directly excited by the actuator.

Figure 8 (a) compares the normalized $\bar{\boldsymbol{M}}$ of each cylinder (from the second to the tenth cylinder) in the first frequency bandgap of the six homogeneous chains, from the 50 mm cylinder chain to the 100 mm cylinder chain. The wide transparent bar represents the numerical analysis results based on the DEM, while the narrow bar represents the experiment data. The experiment data exhibit overall larger magnitudes and show more variations among the chains when compared to the numerical results. This disparity can be attributed to the presence of noise and experimental error in the testing process. Nonetheless, there is notable agreement between the two sets of data, and it is apparent that the magnitude diminishes exponentially across the spatial domain, consistent with the analytical solution described earlier. Additionally, we note that as the length of the cylinder decreases within the chain (leading to a shift of the bandgap frequency towards higher frequencies), the attenuation becomes more pronounced in the spatial domain. Specifically, the attenuation is notably more pronounced in the second bandgap (Fig. 8(b)) compared to that in the first bandgap (Fig. 8(a)). This can be explained through the concept of the effective mass (Eq.(23)) which is defined as a function of frequency (ω). As the wave frequency (ω) approaches a resonance frequency ($\omega_j$), the magnitude of the effective mass undergoes a notable increase. When the magnitude of the effective mass surpasses $4\beta_r/\omega^2$, the wave starts to attenuate. (We can prove this with the dispersion relation in Eq. (31)). Furthermore, the attenuation becomes more pronounced as the effective mass increases. The impact of this effect becomes more prominent at higher frequencies, primarily due to the contribution of the second term in Equation (23), leading to more substantial wave attenuation. Upon estimating $\bar{\beta}$ for the first and second bandgap frequencies in each chain by fitting $e^{-\bar{\beta}(\frac{x}{a})}$ using the data presented in Figs. 8(a) and 8(b) (see Table1 in Appendix C), we observe that $\bar{\beta}$ in the second bandgap is more than three times larger than that in the first bandgap in every chain.

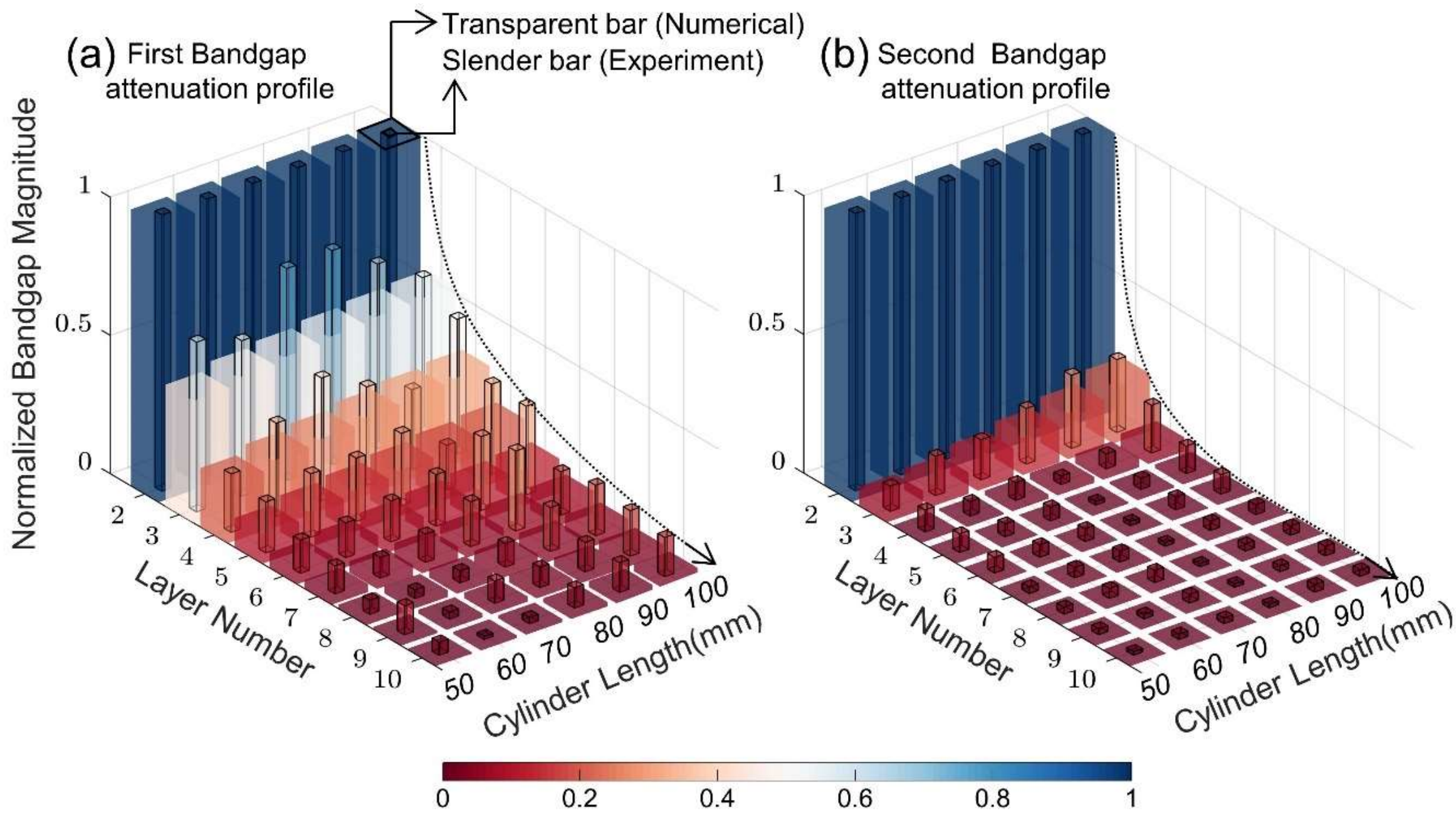

**Figure 8.** Normalized and averaged magnitudes of the frequency response for each cylinder in the first bandgap (a) and second bandgap (b) compared across the six homogeneous chains with varying cylinder lengths. The numerical results obtained from the DEM are represented by the large transparent bars, while the experimental data are depicted by the narrow bars.

## 5.2 Superposition of bandgaps in the graded cylinder chain

In this section, we discuss the efficient method of widening the frequency bandgap (i.e., bandgap engineering) in the cylinder chain. One intuitive approach is to superimpose different bandgaps; for instance, by stacking the four homogeneous cylinder chains with different lengths (ranging from 50 to 80 mm), the bandgap frequencies exhibited in each cylinder chain, as shown in Fig. 7, combine to form a wide bandgap.

Considering the exponential attenuation of waves within the frequency bandgap, we can achieve a wide bandgap using a small number of cylinders. In the homogenous chain, we observe that the wave amplitude is reduced to approximately 80% within the first three cylinders in the first bandgap (Fig. 8(a)). Moreover, the wave attenuation is even more rapid in the second bandgap; more than 80% reduction occurs within the first two cylinders. This implies that a sufficient bandgap can be obtained by employing only two or three cylinders in each length. Hence, constructing a cylinder chain with varying cylinder length and superimposing bandgaps in each elements enable us to achieve a wide bandgap. We also confirm this by optimizing chain with a genetic algorithm (see Appendix E).

We investigate the frequency response of a representative graded chain; the chain length varies linearly from 100 mm to 50 mm with a 2 mm length change, as shown in Fig. 1(test model) and Fig. 6 (FE model). The FRFs of the graded cylinder chain measured from the experiment and calculated with the FEM and DEM are compared in Fig. 9. In the DEM simulation, we directly obtain the wave transmission through numerical integration via Eqs. (32)-(33) under frequency sweep excitations and use fast Fourier transformation (FFT) for the FRF. The FRF calculated with the DEM is in good agreement with those of the experiment and FEM despite the fact that it is a highly simplified discretization model.

We look into the FRFs within 50 kHz, wherein the constituent cylinders of the homogeneous chains exhibit multiple bending vibration modes, leading to the formation of multiple bandgaps. The 100 mm cylinder model has frequency bandgaps at 1.59 ~2.54 kHz (first bandgap), 9.18 ~13.34 kHz (second bandgap), and 22.25 ~31.53 kHz (third bandgap). As the length of the cylinders gradually decreases to 50 mm, the bandgaps expand and shift towards higher frequencies up to 6.07 ~ 9.97 kHz (first bandgap) and 24.87 ~ 48.98 kHz (second bandgap). In the graded chain, where only one cylinder of each length is stacked, we can infer that these bandgaps weakly overlap; the weak bandgaps (stemming from the presence of one cylinder at each length) with gradually shifted frequency range (due to the gradual change of the cylinder length) become overlap.

It is notable that the first bandgap in the homogeneous chain shows significant attenuation only within a narrow frequency range near the beginning of the bandgap (see FEM & DEM results in Fig. 7). As a result, we only observe a partial bandgap around 1~ 6 kHz, even though the overall overlapping range of the bandgap is 1.6 ~ 9 kHz (see Figs. 9(b) and (c)). In the experimental data (Fig. 9(a)), the bandgap appears less distinct, presumably due to damping effects in the test model and minor buckling of the chain during the experiment.

In Figs. 9(a), (b), and (c), a wide frequency bandgap starts to emerge at approximately 9 kHz. This occurs because the evident second bandgap begins to overlap, originating from the 100 mm cylinder; the second bandgap is much clearer (see Fig. 8(b)) and broader (see Fig. 7) compared to the first bandgap. Surprisingly, this bandgap is unbounded given that the strong bandgaps in the 100 mm ~ 50 mm cylinder chains are consecutively superimposed. Moreover, as the frequency becomes higher in the homogeneous chains, the bandgap becomes broader and stronger, while the passband becomes narrower (see Figs. 5(a), 5(b) and, Fig.7). This enhances the strong superposition of the bandgaps at higher frequencies. It's observed that the amplitude of the Frequency Response Function (FRF) within the bandgap region of the experimental data (Fig. 9(a)) surpasses that of the numerical simulations (Figs. 9(b) and 9(c)). This divergence can be attributed to the inclusion of noise, with a noise level estimated at approximately -50 dB.

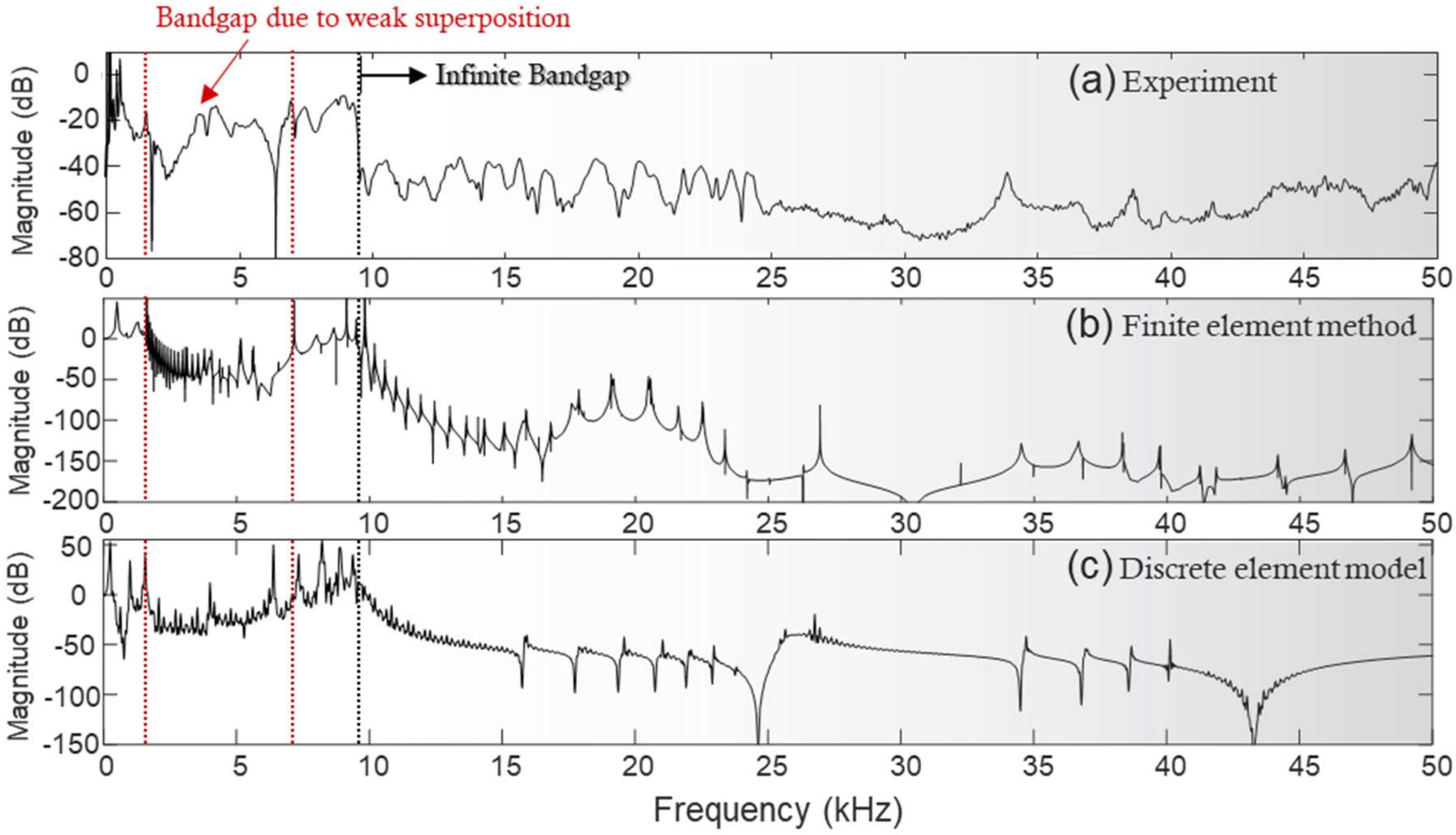

**Figure 9.** Comparison of the frequency response function (FRF) of a one-dimensional graded woodpile lattice. (a) Experimental results (b) Numerical results obtained from the finite element method (FEM). (c) Numerical results obtained from discrete element model (DEM) under linear sweep signal excitation.

To understand the development of the frequency bandgap within the graded cylinder chain, we compare the frequency responses of each cylinder in Figs. 10(b) ~ 10(d), where the color scale indicates the magnitude of the frequency response in dB. The experimental data (Fig. 10(b)) are obtained by calculating the FFT of the temporal responses measured in each cylinder. Similarly, the DEM results (Fig. 10(d)) are obtained through temporal simulations and by applying FFT to the resulting data. The FEM data in Fig. 10(c), on the other hand, are obtained through a steady-state analysis. Notably, all three sets of data exhibit excellent agreement with each other and represent the evolution of the frequency bandgap inside the chain.

We observe that the local resonances of each cylinder prohibit the propagating waves with specific frequencies, and their superposition develops broad frequency bandgaps (see the lines with circles in Fig.10(c)), as explained in section 3. The first local resonance frequencies of the cylinders are within the range of 2.3 kHz - 5.9 kHz (blue line with circles in Fig. 10(c)), and the resonance frequency gradually increases as cylinder length decreases (i.e., as the particle index increases). However, in this frequency range, the superposition of the local resonance bandgaps is not sufficient to create a strong bandgap. On the other hand, the second (around 10 ~ 25 kHz) and the third (around 30 ~ 40 kHz) local resonances (red and yellow lines with circles in Fig. 10(c), respectively) produce a strong bandgap through their superposition.

The blocking mechanism of the propagating wave, resulting from the local resonance, is particularly evident in the DEM simulation; at the local resonance frequencies, we observe that the primary mass ($\widetilde{M}$) remains stationary while specific added masses ($m_i$) vibrate out-of-phase with the incoming wave. As a result, the primary mass is dynamically in a force equilibrium state (see Fig.10(a) top). This phenomenon is commonly referred to as *vibration isolation* in classical vibration theory, which is highly effective in blocking vibrations with a specific frequency [29]. However, conventional vibration isolation systems typically have a narrow working frequency range. Our graded cylinder chain, however, efficiently manages and isolates the propagating waves over a broad range of frequencies by leveraging multiple vibration modes of beam bending (see Fig. 10(a) bottom). It's noteworthy that numerous intriguing wave dynamics also emerge when the bandgaps overlap within the graded system. These dynamics encompass phenomena such as wave amplification [6,34,35], boomerang-like wave reflection [6], asymmetric wave transmission in the nonlinear regime [6], Bloch oscillation [36], and more. The presence of multiple local resonances within the graded system enhances the richness of these phenomena across various frequencies. The suggested physics-informed Discrete Element Method (DEM) empowers us to efficiently investigate these wave dynamics using a system with fewer degrees of freedom compared to the conventional finite element method.

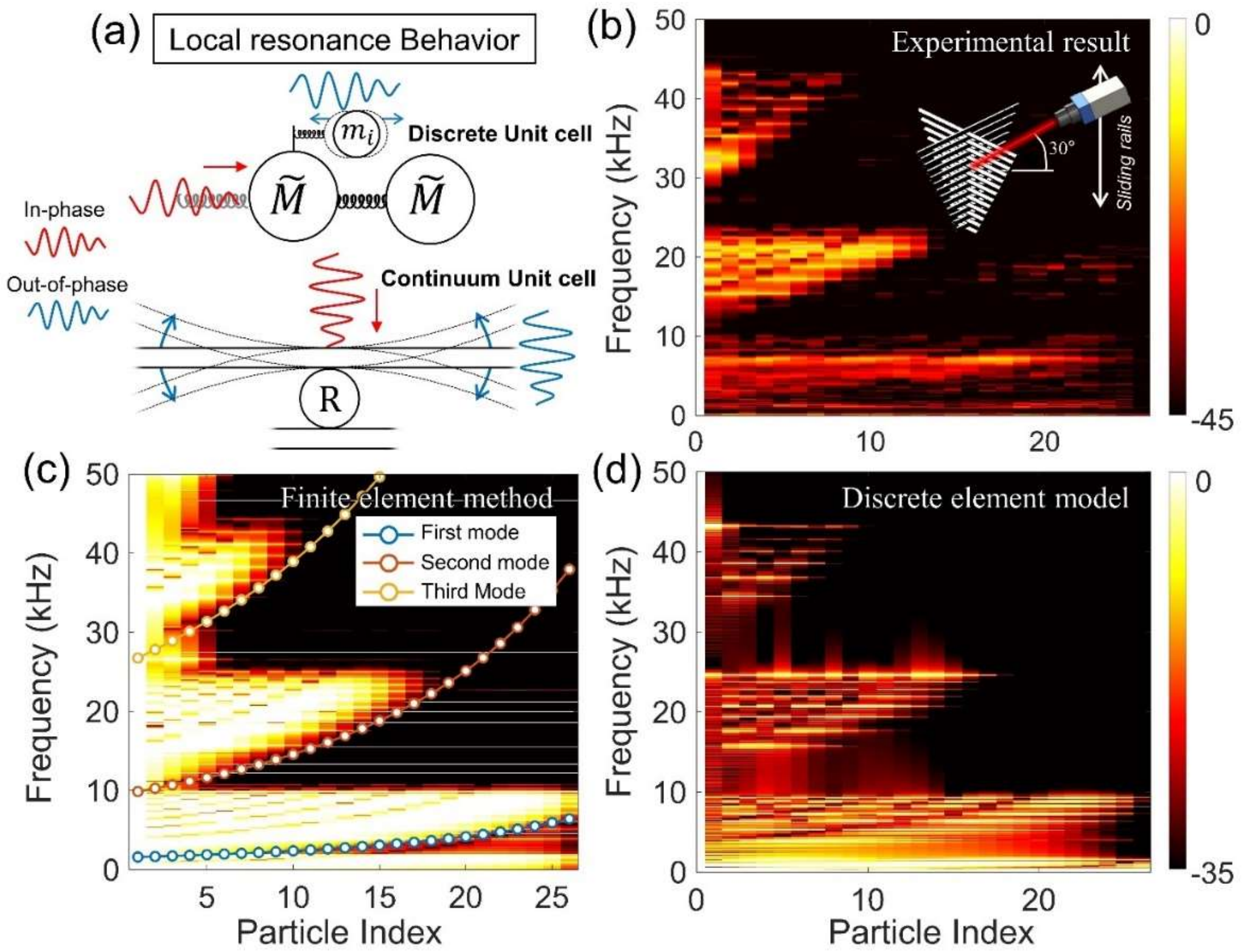

**Figure 10.** (a) Schematics of vibration isolation due to local resonance in DEM (top) and the corresponding vibration of a cylinder beam (bottom). Image maps of the frequency spectrum of the graded cylinder chain (b) measured from experimental and simulation results from (c) FEM and (d) DEM. The inset in (b) illustrates the measurement of each cylinder's vibration using the LDV in the experiment. Also, the lines with circles in (c) represent the (odd-numbered, symmetric) local resonance frequencies in each cylinder obtained from analytic beam theory.

## 6 Conclusion

In this study, we develop an efficient method for constructing a discrete element model (DEM) capable of accurately simulating the oscillations of a continuum beam. This model employs a mass-with-masses system, where a primary mass is connected to multiple auxiliary spring-mass units, to account for multiple vibration modes of the continuum beam. The construction of the DEM is based on the continuum beam theory, specifically the Timoshenko beam theory, and the wave dynamics of a periodic structure. This physics-informed method enables us to construct the DEM analytically. Furthermore, this approach allows for virtually unlimited extension of the DEM in terms of the number of vibration modes considered and the frequency range addressed. We validate the DEMs generated using the proposed method by comparing them with both FEM simulations and experimental data. Through this validation process, we confirm that the DEM is efficient and accurately represents the intricate wave dynamics exhibited by complex woodpile structures.

We also explore the potential of bandgap engineering through the utilization of cylinder chains and the developed DEM method. We find that the frequency bandgap is efficiently tuned by gradually varying the cylinder lengths in the cylinder chain. By leveraging such versatile frequency band structures inside the graded chain, we demonstrate the possibility of constructing an infinite bandgap by consecutively superposing multiple bandgaps in the cylinders. In this graded chain, we verify the so-called vibration isolation effect, by which the propagating wave is efficiently blocked due to the local resonance of the cylinders. We conclude that simplifying a continuum structure into a DEM can be useful for efficiently investigating and designing complex vibration systems. Also, the design approach of broadening the frequency bandgap can be widely used to develop a vibration filter or to suppress vibration/noise in engineering applications.

## Acknowledgment

This research was supported by the National Research Foundation of Korea (NRF) grant funded by the Korean government (MSIT) No. 2020R1A2C2013414. Also, E.K. acknowledges the support by research funds of Jeonbuk National University in 2022. J. Y. is grateful for the support from SNU-IAMD, SNU-IOER, and National Research Foundation grants funded by the Korea government [2023R1A2C2003705 and 2022H1D3A2A03096579 (Brain Pool Plus by the Ministry of Science and ICT)]. J.R. acknowledges the POSCO-POSTECH-RIST Convergence Research Center program funded by POSCO.

## Appendix A. Effective mass density

In this appendix, we describe the effective mass density $(M_{eff})$ of the mass-with-mass type unit cell. As shown in Fig. 4 (main text), unit cells consist of a primary mass $(\widetilde{M})$ with multiple auxiliary masses $(m_N)$ connected by linear springs $(k_N)$. The subscript N denotes the number of vibration modes taken into account. The equation of motion for the $i^{th}$ unit cell and the $j^{th}$ auxiliary mass are as follows, respectively,

$$\widetilde{M}_i \ddot{u}_i = \beta_r \left[u_{i-1} - u_i\right] - \beta_r \left[u_i - u_{i+1}\right] + \sum_{j=1}^{N} k_{i,j} \left(v_{i,j} - u_i\right), \tag{A.1}$$

$$m_{i,j} \ddot{v}_{i,j} = k_{i,j} \left(u_i - v_{i,j}\right). \tag{A.2}$$

Assuming harmonic motion in an infinite space, we can impose $u(v)_i = A_{u(v)} e^{i(\omega t - kx)}$, where $\omega$ and $k$ represent the angular frequency and wavenumber, respectively. The $A_{u(v)}$ is the amplitude of the primary and auxiliary masses. By substituting these relations into Eq. (A.2), we get the following relation,

$$v_{i,j} = \frac{\omega_j^2}{\omega_j^2 - \omega^2} u_i, \tag{A.3}$$

where $\omega_j = \sqrt{k_{i,j}/m_{i,j}}$ is the local resonant frequencies of $j^{th}$ auxiliary spring-mass set. The index *i* is omitted since all the cylinders in the homogeneous chains are identical. By substituting Eq. (A.3) into Eq. (A.1) to eliminate $v_{i,j}$, we get the following expression:

$$M_{eff}(\omega)\omega^2 u_i = \beta_r \left[u_{i-1} - 2u_i + u_{i+1}\right], \tag{A.4}$$

where

$$M_{eff}(\omega) = \widetilde{M} + \sum_{j=1}^{N} m_j \frac{\omega_j^2}{\omega_j^2 - \omega^2}. \tag{A.5}$$

Equation (A.4) represents the equation of motion for conventional monoatomic chains; however, its effective mass density is dependent on the excitation frequency $\omega$.

# Appendix B. Steady state-space response

Here, we elaborate on the calculation of the frequency response function (FRF) for the finite woodpile chain, expressed as a discrete element model. We employ a state-space approach to describe the dynamics of a finite chain, enabling us to obtain a steady-state response as a function of input frequencies without the need for initial conditions. The governing equations of motion (Eqs. (32) and (33)) for the finite chain can be compactly expressed in a general matrix form as follows:

$$\dot{\boldsymbol{q}} = \boldsymbol{A}\boldsymbol{q} + \boldsymbol{B}F_d,$$
$$F_N = \boldsymbol{C}\boldsymbol{q} + \boldsymbol{D}F_d. \tag{B.1}$$

Here, the state vector $\boldsymbol{q}$ comprises the displacement and velocity components of each unit cell. The matrices $\boldsymbol{A}$, $\boldsymbol{B}$, and $\boldsymbol{C}$ corresponds to the state, input, and output matrices, respectively. The direct transformation matrix $\boldsymbol{D}$ is a scalar with a zero value, as no direct input influences the output. The input corresponds to the dynamic disturbance applied to the first particle in the chain, while the output denoted as $F_N = \beta_w u_n$, represents the transmitted contact force at the opposite boundary of the chain. Thus, $\boldsymbol{q}$, $\boldsymbol{A}$, $\boldsymbol{B}$, and $\boldsymbol{C}$ are expressed as:

$$\mathbf{A} = \begin{pmatrix} \mathbf{0} & \boldsymbol{I} \\ \boldsymbol{M}^{-1}\boldsymbol{K} & \mathbf{0} \end{pmatrix}, \quad \boldsymbol{q} = \begin{pmatrix} u_1 \\ \vdots \\ u_P \\ v_{1,1} \\ \vdots \\ v_{P,1} \\ \vdots \\ v_{1,j} \\ \vdots \\ v_{P,j} \\ \dot{u}_1 \\ \vdots \\ \vdots \\ \dot{v}_{P,j} \end{pmatrix}, \quad \boldsymbol{B} = \begin{pmatrix} 0 \\ \vdots \\ 0 \\ 0 \\ \vdots \\ 0 \\ \vdots \\ 0 \\ \vdots \\ 0 \\ 1/\widetilde{M}_1 \\ \vdots \\ \vdots \\ 0 \end{pmatrix}, \quad \boldsymbol{C} = \begin{pmatrix} 0 \\ \vdots \\ \beta_w \\ 0 \\ \vdots \\ 0 \\ \vdots \\ 0 \\ \vdots \\ 0 \\ 0 \\ \vdots \\ \vdots \\ 0 \end{pmatrix}^T, \tag{B.2}$$

where $\boldsymbol{M}$ and $\boldsymbol{K}$ are mass and stiffness matrices determined from the equations of motion in Eqs. (32) and (33), and $\boldsymbol{I}$ is an identity matrix. The mass and stiffness matrices have the following forms:

$$\boldsymbol{M} = \begin{pmatrix} \widetilde{\boldsymbol{M}}_1 & & & & & & & & 0 \\ & \ddots & & & & & & & \\ & & \widetilde{\boldsymbol{M}}_P & & & & & & \\ & & & \boldsymbol{m}_{1,1} & & & & & \\ & & & & \ddots & & & & \\ & & & & & \boldsymbol{m}_{P,1} & & & \\ & & & & & & \ddots & & \\ & & & & & & & \boldsymbol{m}_{1,j} & \\ & & & & & & & & \ddots \\ 0 & & & & & & & & \boldsymbol{m}_{P,j} \end{pmatrix}, \tag{B.3}$$

$$\boldsymbol{K} = \begin{pmatrix} \underline{\boldsymbol{K_{11}}} & \underline{\boldsymbol{K_{12}}} & \cdots & \underline{\boldsymbol{K_{1j}}} \\ \underline{\boldsymbol{K_{21}}} & \underline{\boldsymbol{K_{22}}} & \cdots & \underline{\boldsymbol{K_{2j}}} \\ \vdots & \vdots & \ddots & \vdots \\ \underline{\boldsymbol{K_{j1}}} & \underline{\boldsymbol{K_{j2}}} & \cdots & \underline{\boldsymbol{K_{jj}}} \end{pmatrix}. \tag{B.4}$$

Here, $\boldsymbol{M}$ is a diagonal matrix having element $\widetilde{M}_i$ and $m_{i,j}$, and $P$ is the number cylinder in the chain. $\boldsymbol{K}$ is a symmetric matrix. $\underline{\boldsymbol{K_{11}}}$ is defined as follows:

$$\underline{\boldsymbol{K_{11}}} = \begin{pmatrix} -\boldsymbol{\beta_r} - \boldsymbol{\beta_w} - \sum_{j=1}^{N} \boldsymbol{k_{1,j}} & \boldsymbol{\beta_r} & & & & \\ \boldsymbol{\beta_r} & -2\boldsymbol{\beta_c} - \sum_{j=1}^{N} \boldsymbol{k_{2,j}} & \boldsymbol{\beta_r} & & \boldsymbol{0} & \\ & \boldsymbol{\beta_r} & \ddots & \ddots & & \\ & & \ddots & & \boldsymbol{\beta_r} & \\ & \boldsymbol{0} & & \boldsymbol{\beta_r} & -2\boldsymbol{\beta_c} - \sum_{j=1}^{N} \boldsymbol{k_{P-1,j}} & \boldsymbol{\beta_r} \\ & & & & \boldsymbol{\beta_r} & -\boldsymbol{\beta_r} - \boldsymbol{\beta_w} - \sum_{j=1}^{N} \boldsymbol{k_{P,j}} \end{pmatrix}. \tag{B. 5}$$

$\boldsymbol{K_{1j}}$, $\boldsymbol{K_{j1}}$, and $\boldsymbol{K_{ij}}$ $\forall \boldsymbol{i} = \boldsymbol{j}$ $(\boldsymbol{i}, \boldsymbol{j} \geq \boldsymbol{2})$ are diagonal matrices having values of $k_{i,j}$, $k_{i,j}$, and $-k_{i,j}$. $\boldsymbol{K_{ij}}$ $\forall \boldsymbol{j} \neq \boldsymbol{j}$ $(\boldsymbol{i}, \boldsymbol{j} \geq \boldsymbol{2})$ is null matrix. To calculate the transmission gain, which represents the ratio of the output force to the input force (i.e., $F_N/F_d$) within a specified range of frequency, we solve the Eq. (B.1) by using MATLAB's Bode function.

## Appendix C. Wave Attenuation

Table.1 provides a summary of the average attenuation coefficients obtained from both experimental and numerical analyses for the finite homogeneous chains.

Table 1

The average attenuation coefficients in the 1st & 2nd bandgap frequency region

| cylinder length(mm) | 1st attenuation coefficients ($\bar{\beta}$) | | 2nd attenuation coefficients ($\bar{\beta}$) | |
|---|---|---|---|---|
| | Experiment | Numerical | Experiment | Numerical |
| 50 | 0.57 | 0.78 | 2.22 | 2.69 |
| 60 | 0.53 | 0.70 | 1.87 | 2.56 |
| 70 | 0.44 | 0.66 | 1.82 | 2.23 |
| 80 | 0.42 | 0.62 | 1.62 | 2.00 |
| 90 | 0.39 | 0.58 | 1.33 | 1.79 |
| 100 | 0.31 | 0.56 | 1.08 | 1.56 |

## Appendix D. Euler-Bernoulli beam theory

In this appendix, we explain the Euler-Bernoulli beam theory which is obtained from Eq. (10) in the main text when $\Gamma = \Lambda = 0$. For a linear elastic, isotropic, and homogeneous beam with a constant cross-section, the Euler beam equation is expressed as a single partial differential equation.

$$W_{\xi\xi\xi\xi} - \varphi^4 W = 0, \quad \varphi^4 = \frac{\rho A \omega^2 L^4}{EI}. \tag{D.1}$$

Here we introduce non-dimensional variable $\xi = x/L$. By doing so, a general solution to the differential equation (D.1) can be written as:

$$W(\xi) = C_1 \cosh(\varphi\xi) + C_2 \sinh(\varphi\xi) + C_3 \cos(\varphi\xi) + C_4 \sin(\varphi\xi). \tag{D.2}$$

Similar to the analytic approaches described in section 3.2 of the manuscript, we apply specific boundary conditions associated with the three wavenumbers, assuming an infinite periodic chain (i.e., $\kappa a = 0, \pi,$ and $-i\infty$).

In the case where $\kappa a = 0$, we can apply sliding boundary conditions at the left end of the half beam (center of the entire beam) and free boundary conditions at the right end of the half beam, which give rise to the following conditions:

$$\text{Sliding BC: } W_{\xi}\big]_{\xi=0} = 0, \; W_{\xi\xi\xi}\big]_{\xi=0} = 0, \tag{D.3}$$

$$\text{Free BC: } W_{\xi\xi}\big]_{\xi=1} = 0, \; W_{\xi\xi\xi}\big]_{\xi=1} = 0. \tag{D.4}$$

In the case where $\kappa a = \pi$, we can apply an elastic support with contact stiffness at the left of the half beam, leading to the following conditions:

$$\text{Elastic support: } W_{\xi\xi\xi}\big]_{\xi=0} = 2\eta_e W\big]_{\xi=0}, \; W_{\xi}\big]_{\xi=0} = 0, \tag{D.5}$$

where $\eta_e$ is expressed as $\beta_r L^3 / EI$, representing the dimensionless linearized contact stiffness for the Euler beam. The free boundary condition at the right end of the half beam is the same as (D.4).

In the case where $\kappa a = -i\infty$, we apply a fixed boundary condition at the left of the half beam:

$$\text{Fixed BC: } W\big]_{\xi=0} = 0, \; W_{\xi}\big]_{\xi=0} = 0. \tag{D.6}$$

Additionally, free boundary conditions are employed at the right end of the half beam as described in (D.4). Finally, by substituting each case of boundary condition into equation (D.2), we obtain the characteristic equation.

$$\text{At } \kappa a = 0, \left[\tan\varphi + \tanh\varphi\right] = 0, \tag{D.7}$$

At $\kappa a = \pi$,

$$-2\eta_e\left[\cos\varphi\cosh\varphi + 1\right] + \varphi^3\left[\cosh\varphi\sin\varphi + \cos\varphi\sinh\varphi\right] = 0, \tag{D.8}$$

$$\text{At } \kappa a = -i\infty, \left[\cos\varphi\cosh\varphi + 1\right] = 0. \tag{D.9}$$

In Eq. (D.7), we obtain the onset frequency $(\omega_{O_i})$, while in Eq. (D.8) we obtain the cutoff frequency $(\omega_{C_i})$ of each passband. Furthermore, Eq. (D.9) gives the cutoff frequency $(\omega_{C_i})$ within each bandgap. It is interesting to note that the characteristic equation (D.8) determining the cutoff frequency is mathematically expressed as a combination of two terms in (D.7) and (D.8), which determine the onset frequency and local resonance frequency, respectively.

## Appendix E. Optimized chain

In this section, we explore the formation of an optimized cylinder chain with wide frequency bandgaps using the DEM. For the optimization process, we employ a Breether Genetic Algorithm (BGA) (Kim et al., 2017). The BGA is configured to select 26 cylinders, from top to bottom, from a pool of cylinders with varying lengths; the length of the cylinders ranges from 100 mm to 50 mm with 2 mm intervals. The objective of the optimization is to design a chain that exhibits wide frequency bandgaps within the range of 50 kHz. To achieve this, we define an objective function that involves integrating the magnitude of FRFs within the target frequency ($\omega_T$ = 50 kHz). We multiply a weight function $(1 - \omega/\omega_T)$ to the FRF, giving more weight at lower frequencies to get wider bandgap in lower frequency. Then, we minimize the objective function, which is expressed as $\boldsymbol{min}[\int_0^{\omega_T} FRF(\omega) \times (1 - \omega/\omega_T)\, d\omega]$ . It is worth noting that creating wide bandgaps at low frequencies poses more challenges compared to creating wide bandgaps at higher frequencies in a local resonance system. We observe that the width of bandgap reduces as the frequency decreases in homogeneous cylinder chain (see Fig.7). We can confirm that the optimized chain slightly changes depending on the weight function, however, their FRFs are almost identical. We compare three chain models: (1) the optimized chain, (2) the sorted chain with cylinder lengths in descending order from the optimized chain, (3) a graded cylinder chain as follows:

(1) Optimized chain

[92, 100, 54, 66, 96, 68, 86, 82, 86, 72, 66, 98, 66, 68, 74, 52, 56, 82, 62, 82, 78, 56, 92, 62, 88, 56]

(2) Sorted chain

[100, 98, 96, 92, 92, 88, 86, 86, 82, 82, 82, 78, 74, 72, 68, 68, 66, 66, 66, 62, 62, 56, 56, 56, 54, 52]

(3) Graded chain

[100, 98, 96, 94, 92, 90, 88, 86, 84, 82, 80, 78, 76, 74, 72, 70, 68, 66, 64, 62, 60, 58, 56, 54, 52, 50]

Here the number represents the cylinder length in mm.

Figure S1(a) compares the frequency responses for the three chains. We observe that the FRFs of the three chains are almost identical among each other. We also compare the FRFs at each cylinder in the three chains to look into the wave propagation inside of the chains, as shown in Fig. S1(b). The sorted and graded chains show similar internal wave dynamics; the propagating waves are blocked by the gradually varying local resonances in the cylinder chain. In the optimized chain, the bandgap in the low-frequency regime (around 1 kHz~ 6kHz) is less prominent, while the wide bandgap at high frequency ( >9 kHz) forms earlier (within the first ten particles) compared to the other chains.

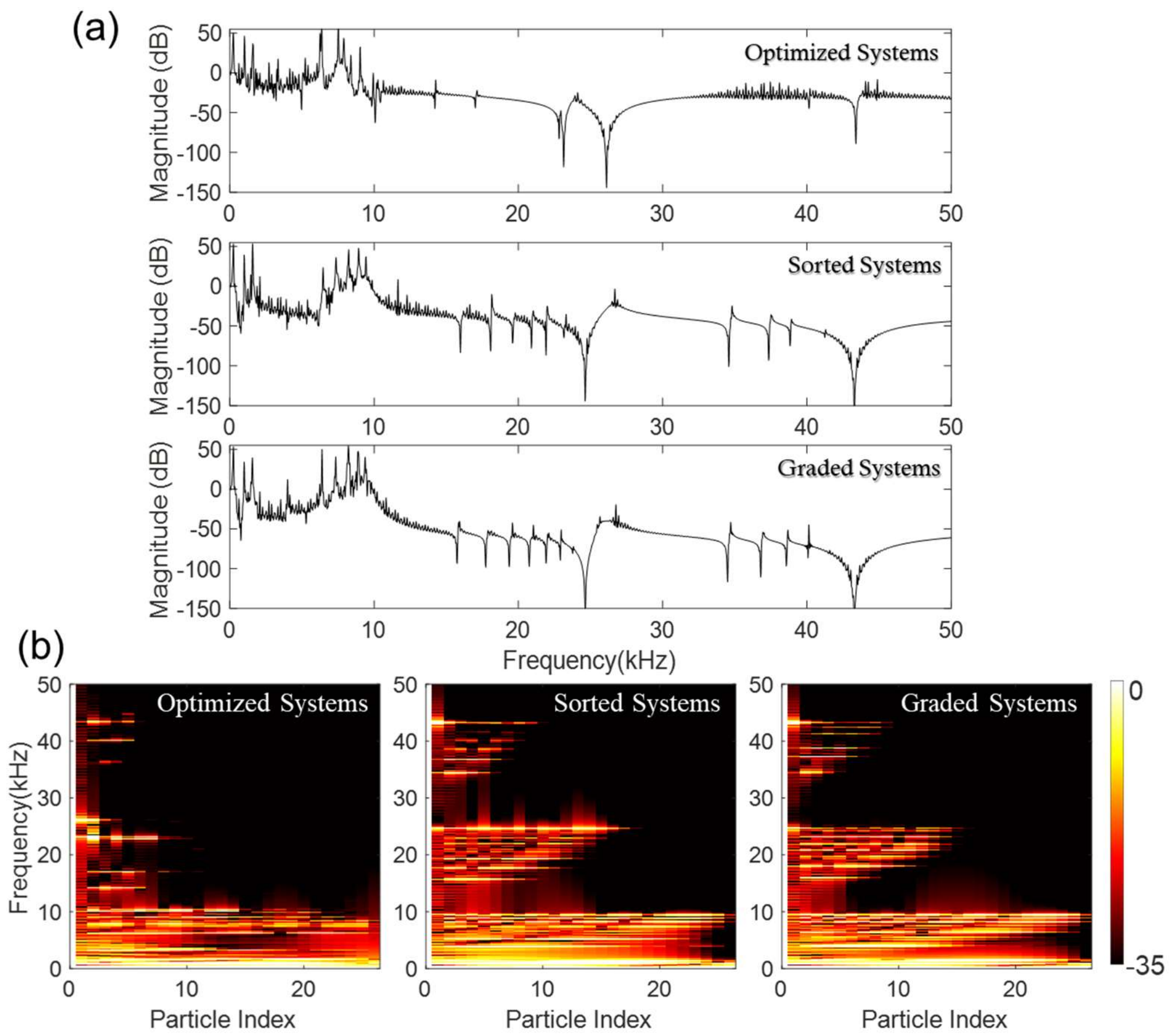

**Figure S1** (a) Frequency responses of the optimized chain (top), the sorted chain (middle), and the graded chain (bottom). (b) Image maps of the frequency responses in the optimized chain (left), sorted chain (middle), and graded chain (right).